\documentclass[11pt]{article}
\usepackage{cite}
\usepackage{mdframed}
\usepackage{epsfig,subfigure}
\usepackage{amssymb}
\usepackage[fleqn]{amsmath}
\usepackage{color}
\usepackage{arydshln}
\newcommand{\define}{\stackrel{\triangle}{=}}
\def\QED{\mbox{\rule[0pt]{1.5ex}{1.5ex}}}

\usepackage{graphicx}
\usepackage{color}
\usepackage[dvipsnames]{xcolor}

\definecolor{armygreen}{rgb}{0.29, 0.33, 0.13}

\usepackage{amssymb}
\usepackage{amsmath,amsfonts,amssymb}
\usepackage{verbatim}
\usepackage{stfloats}
\usepackage[bookmarks=false]{}
\usepackage{algorithm}
\usepackage{algcompatible}

\newtheorem{theorem}{Theorem}
\newtheorem{corollary}{Corollary}

\newtheorem{lemma}{Lemma}

\newtheorem{example}{Example}
\newcommand\blfootnote[1]{%
  \begingroup
  \renewcommand\thefootnote{}\footnote{#1}%
  \addtocounter{footnote}{-1}%
  \endgroup
}

\begin{document}
\date{}
%\title{%On Multiple Anonymous Communication Sessions
%The Linear Capacity of Anonymous \\
%Communications %\\
%with Multiple %Communication 
%Sessions
%%Anonymous Communications over a Multiple Access Channel
%%\thanks{This work is supported by NSF grants CCF-1317351 and CCF-0963925.}
%}
%\author{\IEEEauthorblockN{Hua Sun}
%\IEEEauthorblockA{Department of Electrical Engineering\\
%University of North Texas, Denton, TX 76203\\
%hua.sun@unt.edu}
%}

\title{
Private Information Delivery
%The Capacity of Anonymous Communications
%\thanks{This work is supported by NSF grants CCF-1317351 and CCF-0963925.}
}
\author{\normalsize Hua Sun \\
}

\maketitle

\blfootnote{Hua Sun (email: hua.sun@unt.edu) is with the Department of Electrical Engineering at the University of North Texas.}
\maketitle

\begin{abstract}
We introduce the problem of private information delivery (PID), comprised of $K$ messages, a user, and $N$ servers (each holds $M\leq K$ messages) that wish to deliver one out of $K$ messages to the user privately, i.e., without revealing the delivered message index to the user. %This AID problem may be viewed as the dual of the private information retrieval problem. 
The information theoretic capacity of PID, $C$, is defined as the maximum number of bits of the desired message that can be privately delivered per bit of total communication to the user. For the PID problem with $K$ messages, $N$ servers, $M$ messages stored per server, and $N \geq \lceil \frac{K}{M} \rceil$, we provide an achievable scheme of rate $1/\lceil \frac{K}{M} \rceil$ and an information theoretic converse of rate $M/K$, i.e., the PID capacity satisfies $1/\lceil \frac{K}{M} \rceil \leq C \leq M/K$. This settles the capacity of PID when $\frac{K}{M}$ is an integer.
When $\frac{K}{M}$ is not an integer, we show that the converse rate of $M/K$ is achievable if $N \geq \frac{K}{\gcd(K,M)} - (\frac{M}{\gcd(K,M)}-1)(\lfloor \frac{K}{M} \rfloor -1)$, and the achievable rate of $1/\lceil \frac{K}{M} \rceil$ is optimal if $N = \lceil \frac{K}{M} \rceil$.
Otherwise if $\lceil \frac{K}{M} \rceil < N < \frac{K}{\gcd(K,M)} - (\frac{M}{\gcd(K,M)}-1)(\lfloor \frac{K}{M} \rfloor -1)$, we give an improved achievable scheme and prove its optimality for several small settings.
%is $M/K$ when $N \geq K$, and $1/{\lceil \frac{K}{M} \rceil}$ when $\lceil \frac{K}{M} \rceil \leq N < \frac{K}{\gcd(K, M)}$.
\end{abstract}

\newpage

\allowdisplaybreaks
\section{Introduction}
{\color{black}Consider a dataset comprised of $K$ identically distributed messages and stored over $N$ servers. The servers wish to deliver one of the messages to a user without revealing the identity of the message delivered, i.e., the user does not know which message is delivered to him. For example, the dataset may be medical records from a hospital and each message represents the medical record of a patient. The hospital would like to send the medical record of a patient externally (e.g., for analysis of certain disease that goes beyond the capability of the current hospital), and it is desirable that the name of the patient is not revealed (i.e., the privacy of the patient is preserved). For another example, suppose a company outsources some of its user activity log data externally for statistical analysis, while it does not wish to reveal sensitive information about the user identities (e.g., names, addresses, groups).
We call this problem private\footnote{{\color{black}In a previous version of this work \cite{AID}, the problem is called \emph{anonymous} information delivery. We make a clear distinction of privacy and anonymity here, where privacy refers to the behavior or interest of an entity (e.g., which message is delivered) and anonymity refers to the entities of certain activity (e.g., who pays the bill \cite{Chaum_DC}). %We would like to thank Professor Syed Jafar for this suggestion (if you do not like this, let me know).
}} information delivery (PID). %In this work, we consider the elemental case where one out of the $K$ messages is delivered privately.
}

%In the private information retrieval (PIR) problem \cite{pirfirst}, a user wishes to retrieve one out of $K$ messages from $N$ replicated servers, without revealing which message he wants to each individual server. The motivation here is to protect the privacy of the user so that the user may obtain desired information while hiding his preference. In this work, we switch the focus to the privacy of the dataset at the servers - the $N$ servers wish to deliver one out of $K$ messages to the user anonymously, i.e., the user does not know which message is delivered to him. We call this problem anonymous information delivery (AID). This AID problem models the situations where the delivered message identity needs to be kept perfectly private. For example, the servers might outsource some of their data externally for analysis, but do not want to reveal the identity of the data (e.g., year, location, version number). 
%{\color{blue} For example, the user might request the historical record of one entity for a sample study. For this purpose, any particular entity might work and the servers wish to fully anonymize the record index.} 

This PID problem is trivial for a centralized system, i.e., there is a single server that stores all $K$ messages. In this case, no matter which message the server wishes to deliver, the server simply sends the message to the user and all $K$ choices are indistinguishable from the user. Recently, a fully distributed system is studied in \cite{Sun_Anonymous}, where there are $K$ messages and $N = K$ servers, each stores one message. An example with $K = 3$ and an optimal private coding strategy are shown below. Here we have 3 independent messages $W_1, W_2, W_3$ (one bit each). The servers are equipped with some correlated random variables $z_1, z_2, z_1 + z_2$ that are independent of the messages and $z_1, z_2$ are two i.i.d. fair coin tosses. 
\begin{eqnarray}
\begin{array}{|c|c|c|c|c|}\hline
%&\multicolumn{2}{c}{\mbox{Prob. 1/2}} \vline&\multicolumn{2}{c}{\mbox{Prob. 1/2}} \vline\\ \cline{2-5}
%    & \mbox{Want $W_1 $} &\mbox{Want $W_2 $}& \mbox{Want $W_1$} &\mbox{Want $W_2$} \\
%   \hline		
& \mbox{\small Server 1} & \mbox{\small Server 2} & \mbox{\small Server 3} \\ \hline
\mbox{\small Storage}&W_1, z_1 & W_2, z_2 & W_3, z_1 + z_2\\ \hline
\mbox{\small Answer for $W_1$}& W_1 + z_1 & z_2 & z_1+z_2\\ \hline
\mbox{\small Answer for $W_2$}&  z_1 & W_2 + z_2 & z_1+z_2\\ \hline
\mbox{\small Answer for $W_3$}& z_1 & z_2 & W_3 + z_1+z_2\\ \hline
\end{array}
\label{ach1}
\end{eqnarray}

To ensure information theoretic privacy, we need to guarantee that regardless of the message index delivered, the answers seen by the user are identically distributed and the decoding rule remains the same (otherwise, the decoding rule reveals information about the message delivered). For the scheme above, no matter $W_1$, $W_2$, or $W_3$ is to be delivered, the user sees 3 i.i.d. random bits and to decode the desired message, he always adds up the 3 answering strings. In \cite{Sun_Anonymous}, it is proved that the communication rate of $1/3$ is optimal, where the rate is defined as the number of bits privately delivered per bit of total answers sent to the user. For the above $N = K$ and each server stores $M = 1$ message case, the maximum rate (termed the capacity, $C$) is $1/K$. Further, it is necessary for each server to hold 1 bit of correlated randomness and for all servers to hold $K-1$ bits of correlated randomness, per message bit.

As the fully distributed and centralized cases are well understood, our goal in this paper is to study the intermediate partially distributed case - each server stores $M$ out of $K$ messages ($1 \leq M \leq K$). We are restricted to replicated systems (i.e., we do not allow coded messages or splitting one message to several servers) in this work\footnote{It turns out that the PID problem is trivial when we may distribute (a distinct part of) \emph{each} message to \emph{each} server as in this case, rate 1 can be achieved easily and the system is essentially centralized in the sense of PID. Therefore for the PID problem, the more interesting case of distributed systems refers to that \emph{some} message is not available at all at \emph{some} server, and we wish to confuse the user about which message is delivered. }, as a first step towards more complex scenarios and a practical set-up for distributed storage systems. Note that we allow the design of the $M$ messages stored. That is, we wish to find the best replication strategy and the corresponding private delivery scheme. The main motivation of this work is to characterize the capacity of PID for replicated systems, as a function of the number of messages, $K$, the number of servers, $N$, and the number of messages stored per server, $M$.
%for arbitrary number of messages $K$, arbitrary number of servers $N$, and arbitrary number of messages $M$ stored per server.

As an example, consider the setting where we have $K = 3$ messages, $N = 3$ servers and $M = 2$ messages are stored per server. The storage and correlated randomness design and the private coding scheme are shown below. Here each message is made up of two symbols from $\mathbb{F}_5$, $W_1 = (a_1, a_2)$, $W_2 = (b_1, b_2)$ and $W_3 = (c_1, c_2)$. $z$ is a common random variable shared by the servers and $z$ is uniformly distributed over $\mathbb{F}_5$ (independent of the messages).
\begin{eqnarray}
\begin{array}{|c|c|c|c|c|}\hline
%&\multicolumn{2}{c}{\mbox{Prob. 1/2}} \vline&\multicolumn{2}{c}{\mbox{Prob. 1/2}} \vline\\ \cline{2-5}
%    & \mbox{Want $W_1 $} &\mbox{Want $W_2 $}& \mbox{Want $W_1$} &\mbox{Want $W_2$} \\
%   \hline		
& \mbox{\small Server 1} & \mbox{\small Server 2} & \mbox{\small Server 3} \\ \hline
\mbox{\small Storage}&W_1, W_2, z & W_2, W_3, z & W_3, W_1, z\\ \hline
& & & \\ [-0.6em]
\mbox{\small Answer for $W_1$}& \frac{3}{2}a_1 - \frac{1}{2}a_2 + z & -2z & -\frac{1}{2} a_1 + \frac{1}{2} a_2 + z \\ 
& & & \\  [-0.6em]
\hline
\mbox{\small Answer for $W_2$}&  2b_1 - b_2 + z & -b_1 + b_2 - 2z  & z\\ \hline
\mbox{\small Answer for $W_3$}& z & 3c_1 - c_2 - 2z & -2c_1 + c_2 + z\\ \hline
\end{array}
\label{ach2}
\end{eqnarray}

We denote the answer from Server $n, n \in \{1,2,3\}$ by $A_n$. Note that $A_n$ is a function of the storage at Server $n$. To decode the desired message, in all 3 cases where $W_1, W_2$ or $W_3$ is delivered, the user employs the same decoding strategy, as follows.
\begin{eqnarray}
\mbox{Desired Symbol 1} &=& A_1 + A_2 + A_3 \\
\mbox{Desired Symbol 2} &=& A_1 + 2A_2 + 3A_3 
\end{eqnarray}
Further, in all 3 cases, the user receives 3 uniformly random symbols over $\mathbb{F}_5$, thus perfect privacy is achieved. The rate achieved is $2/3$ as 2 symbols are delivered over 3 answering symbols. As we will show later by an information theoretic converse, the rate of $2/3$ is also the maximum possible. Thus the capacity of PID is $2/3$ in this case. %We are able to fully generalize this result.

The main contribution of this work is summarized next. We first show that $1/\lceil \frac{K}{M} \rceil \leq C \leq M/K$ by an achievable scheme of rate $1/\lceil \frac{K}{M} \rceil$ and a converse of rate $M/K$. As a result, we have $C = M/K$ when $\frac{K}{M} \in \mathbb{Z}$. Otherwise, if $\frac{K}{M} \notin \mathbb{Z}$, we prove that when $N \geq \frac{K}{\gcd(K,M)} - (\frac{M}{\gcd(K,M)}-1)(\lfloor \frac{K}{M} \rfloor - 1)$, the converse rate of $M/K$ is achievable, and when $N = \lceil \frac{K}{M} \rceil$, the achievable rate of $1/\lceil \frac{K}{M} \rceil$ is optimal. For the uncovered regime where $\lceil \frac{K}{M} \rceil < N < \frac{K}{\gcd(K,M)} - (\frac{M}{\gcd(K,M)}-1)(\lfloor \frac{K}{M} \rfloor - 1)$, we provide an improved achievable scheme %of rate $??$ 
and show that it is optimal for certain small cases.  Therefore, we have characterized the capacity of PID for most cases, and provided approximations for remaining cases.

\bigskip

{\it Notation: %$\mathbb{N}$ is the set of natural numbers. 
For  integers $N_1, N_2, N_1 \leq N_2$, define the notation $[N_1:N_2]$ as the set $\{N_1, N_1+ 1,\cdots, N_2\}$ and $(N_1:N_2)$ as the vector $(N_1, N_1 + 1, \cdots, N_2)$. 
For an index set $\mathcal{I} = \{i_1, i_2, \cdots, i_n\}$, the notation $A_{\mathcal{I}}$ represents the set $\{A_{i}: i \in \mathcal{I}\}$. For an index vector $\overrightarrow{\mathcal{I}} = (i_1, i_2, \cdots, i_n)$, the notation $A_{\overrightarrow{\mathcal{I}}}$ represents the vector $(A_{i_1}, A_{i_2}, \cdots, A_{i_n})$. 
For sets (vectors) $\mathcal{I}_1, \mathcal{I}_2$, we define $\mathcal{I}_1/\mathcal{I}_2$ as the set (vectors) of elements that are in $\mathcal{I}_1$ and not in $\mathcal{I}_2$ (in original order). 
The notation $X \sim Y$ is used to indicate that random variables $X$ and $Y$ are identically distributed. For a matrix ${\bf F}$ with $i$ rows and $j$ columns, if we wish to highlight its dimension, we will write ${\bf F}_{i \times j}$. For an index vector $\overrightarrow{\mathcal{I}} = (i_1, i_2, \cdots, i_n)$, the notation ${\bf F}_{[\overrightarrow{\mathcal{I}},:]}$ represents the submatrix of ${\bf F}$ formed by retaining only the rows corresponding to the elements of the vector $\overrightarrow{\mathcal{I}}$. The notation ${\bf F}_{[:, \overrightarrow{\mathcal{I}}]}$ is defined similarly (with respect to the columns).
The notation ${\bf I}_j$ represents the identity matrix with dimension $j \times j$ and the notation ${\bf 0}$ represents a matrix where each element is 0.
%when $A$ is a set, and the length of a tuple when $A$ is a tuple.  
%For $n\in\{1,2\}$ we define $\bar{n}$ as the complement of $n$, i.e., $\bar{n}=1$ if $n=2$ and $\bar{n}=2$ if $n=1$.
}

\section{Problem Statement}\label{sec:model}
Consider $K$ independent messages $W_1, \cdots, W_K$. Each message is comprised of $L$ i.i.d. uniform symbols from a finite field $\mathbb{F}_p$. In $p$-ary units,
\begin{eqnarray}
H(W_{1}) &=& \cdots = H(W_{K}) = L, \label{h1}\\
H(W_{1}, \cdots, W_{K}) &=& H(W_{1}) + \cdots + H(W_{K}). \label{h2}
\end{eqnarray}
There are $N$ servers, and each server stores $M$ out of the $K$ messages. We denote the storage variable at Server $n$ as $S_n$.
\begin{eqnarray}
S_n = W_{\mathcal{S}_n}, ~\mathcal{S}_n \subset [1:K], |\mathcal{S}_n| = M. \label{store}
\end{eqnarray}

The servers share a common random variable $Z$, and $Z$ is independent of the messages. 
\begin{eqnarray}
H(Z, W_1, \cdots, W_K) = H(Z) + H(W_1) + \cdots + H(W_K). \label{zk_ind}
\end{eqnarray}

The servers privately generate $\theta \in [1:K]$ and wish to deliver $W_\theta$ to a user while keeping $\theta$ a secret from the user. Depending on $\theta$, there are $K$ strategies that the servers could employ to privately deliver the desired message. For example, if $\theta = k$, then in order to deliver $W_k$, Server $n \in [1:N]$ sends an answer $A_n^{[k]}$ to the user. The answer $A_n^{[k]}$ is a function of $S_n, Z$,
\begin{eqnarray}
\forall k \in [1:K], n \in [1:N], ~~~ H(A_n^{[k]} | S_n, Z) = 0. \label{det}
\end{eqnarray}
 
From all $N$ answers, the user decodes the desired message with zero error. %with a decoding mapping $g$. Note that the user is not allowed to learn anything about the index of the desired message, so the decoding rule does not depend on $k$. 
%The decoding mapping $g$ is known at every node.
\begin{eqnarray}
H(W_k | A_1^{[k]}, A_2^{[k]}, \cdots, A_N^{[k]}) = 0.
%W_k = g(A_1^{[k]}, A_2^{[k]}, \cdots, A_N^{[k]}).
\label{corr}
\end{eqnarray}
To ensure privacy, the communication strategies must be indistinguishable (identically distributed) from the perspective of the user, i.e., the following privacy constraint must be satisfied, $\forall k\in[1:K]$,
\begin{eqnarray}
&&
\mbox{[Privacy]} ~~~~(A_1^{[1]}, A_2^{[1]}, \cdots, A_N^{[1]}, W_1) \sim (A_1^{[k]}, A_2^{[k]}, \cdots, A_N^{[k]}, W_k).
\label{privacy}
\end{eqnarray}
The privacy constraint (\ref{privacy}) is equivalent to the condition that the answers are i.i.d. and the (deterministic) decoding mappings from the answers to the desired message are the same for all $k$.

%To communicate $W_k$ anonymously, each transmitter uses the channel $N$ times (it is easy to see that the number of channel users does not depend on the desired transmitter index $k$).
The PID \emph{rate} characterizes how many symbols of desired information are delivered per symbol of total delivery, and is defined as 
\begin{eqnarray}
R \triangleq \frac{L}{\sum_{n=1}^N D_n}
\end{eqnarray}
where $D_n$ is the number of symbols sent from Server $n$ to the user.

%Note that by symmetry, the number of channel uses for each transmitter does not depend on the desired transmitter index.
%An AID scheme consists of two parts: 1) storage design of $\mathcal{S}_1, \mathcal{S}_2, \cdots, \mathcal{S}_N$ and 2) answers design of $A_1^{[k]}, A_2^{[k]}, \cdots, A_N^{[k]}$.
A rate $R$ is said to be achievable if there exists a PID scheme of rate greater than or equal to $R$, for which zero error decoding is guaranteed.
The supremum of achievable rates (over all storage design ${S}_1, {S}_2, \cdots, {S}_N$ and all PID schemes) is called the capacity $C$. 
%When the encoding and decoding mappings are restricted to linear schemes (i.e., only linear combination operations are allowed), then the capacity is called the linear capacity. 

%The individual randomness size $\rho$ measures the average amount of correlated randomness at each transmitter relative to the message size.
%(by symmetry, without loss of generality, we assume that each transmitter holds the same amount of correlated randomness, i.e., $H(Z_1) = \cdots = H(Z_K)$). 
The randomness size $\eta$ measures the amount of common randomness at the servers relative to the message size.
\begin{eqnarray}
%\rho &=& \frac{1}{K} \sum_{i=1}^K \frac{H(Z_i)}{L} \label{rho_def}\\
\eta &=& \frac{H(Z)}{L}. \label{eta_def}
\end{eqnarray}
In this work, we focus on the capacity $C$ and allow as much common randomness as needed.

\section{Results}%: Capacity of Anonymous Information Delivery}
In this section, we state the main results of this work. We start with an approximation of the PID capacity, stated in the following theorem.
\begin{theorem}\label{thm:approx}
For the private information delivery problem with $K$ messages, $N \geq \lceil \frac{K}{M} \rceil$ servers and $M$ messages per server, the capacity satisfies
\begin{eqnarray}
1/\lceil \frac{K}{M} \rceil \leq C \leq M/K.
\end{eqnarray}
\end{theorem}

We need $N \geq \lceil \frac{K}{M} \rceil$ because otherwise the total storage available at all servers, $NM$, is smaller than the number of messages, $K$, and we can not guarantee that all messages can be delivered correctly.
To prove Theorem \ref{thm:approx}, we provide an achievable scheme of rate $1/\lceil \frac{K}{M} \rceil$ and a converse of rate $M/K$. The details are presented in Section \ref{sec:approx}.

The bounds in Theorem \ref{thm:approx} match when $\frac{K}{M}$ is an integer. Therefore, in this case, we obtain the exact capacity of PID, stated in the following corollary.
\begin{corollary}\label{cor:int}
For the private information delivery problem with $K$ messages, $N$ servers and $M$ messages per server, if $\frac{K}{M} \in \mathbb{Z}$, $N \geq \frac{K}{M}$, the capacity is $C = M/K$.
\end{corollary}

{\it Remark: When $N = K, M = 1$ (the fully distributed system), $C = 1/K$ and this recovers the capacity result of Theorem 1 in \cite{Sun_Anonymous}. When $M = K$, we have the fully centralized system and $C = 1$.
}

When $\frac{K}{M}$ is not an integer, the bounds in Theorem \ref{thm:approx} represent a reasonable approximation of the PID capacity. The inverse of the capacity ($\frac{1}{C}$, referred to as the optimal download cost) is characterized to within a 1 symbol gap ($= \lceil K/M \rceil - K/M$).

Next, we proceed to consider the conditions on the number of servers $N$ such that the bounds in Theorem \ref{thm:approx} are tight.
\begin{theorem}\label{thm:tight}
For the private information delivery problem with $K$ messages, $N \geq \lceil \frac{K}{M} \rceil$ servers, $M$ messages per server, and $\frac{K}{M} \notin \mathbb{Z}$, 
the converse rate of $M/K$ is achievable if $N \geq \frac{K}{\gcd(K,M)} - (\frac{M}{\gcd(K,M)}-1)(\lfloor \frac{K}{M} \rfloor - 1)$, and the achievable rate of $1/\lceil \frac{K}{M} \rceil$ is optimal if $N = \lceil \frac{K}{M} \rceil$.
\end{theorem}

The proof of Theorem \ref{thm:tight} is presented in Section \ref{sec:tight}.

Combining Theorem \ref{thm:approx} and Theorem \ref{thm:tight}, we have characterized the PID capacity when the number of servers is either small or large. In particular, the full regime is characterized when $M = 2$. This result is stated in the following corollary.
\begin{corollary}
For the private information delivery problem with $K$ messages, %($\frac{K}{2} \notin \mathbb{Z}$), 
$N$ servers and $M = 2$ messages per server, the capacity is
\begin{eqnarray}
C = 
\left\{
\begin{array}{ccl}
2/K && \mbox{when} ~~ N \geq \lceil \frac{K}{2} \rceil + 1 \\
\\ [-1em]
1/{\lceil \frac{K}{2} \rceil} && \mbox{when} ~~ N = \lceil \frac{K}{2} \rceil \\
\\ [-1em]
0 && \mbox{when} ~~ N < \lceil \frac{K}{2} \rceil 
\end{array}
\right.\end{eqnarray}
\end{corollary}

{\it Proof:} The case for even $K$ is obvious (covered in Corollary \ref{cor:int}) and we only need to consider the case when $K$ is odd. 
We prove that $\lceil \frac{K}{2} \rceil + 1 = \frac{K}{\gcd(K,M)} - (\frac{M}{\gcd(K,M)}-1)(\lfloor \frac{K}{M} \rfloor - 1)$, when $M = 2$. Plugging in $M=2$ to the RHS, we have $K - \lfloor \frac{K}{2} \rfloor + 1 = \mbox{LHS}$ and the proof is complete.
\hfill\QED

Note that when $M=2$, the difference between the two thresholds \Big($\lceil \frac{K}{M} \rceil$ and $\frac{K}{\gcd(K,M)} - (\frac{M}{\gcd(K,M)}-1)(\lfloor \frac{K}{M} \rfloor - 1)$\Big) on $N$ is exactly 1. Then we know that the thresholds (conditions for the bounds in Theorem \ref{thm:approx} to be tight) can not be improved in general. %are the best possible in general. 

\begin{figure}[h]
\begin{center}
\includegraphics[width=3.5 in]{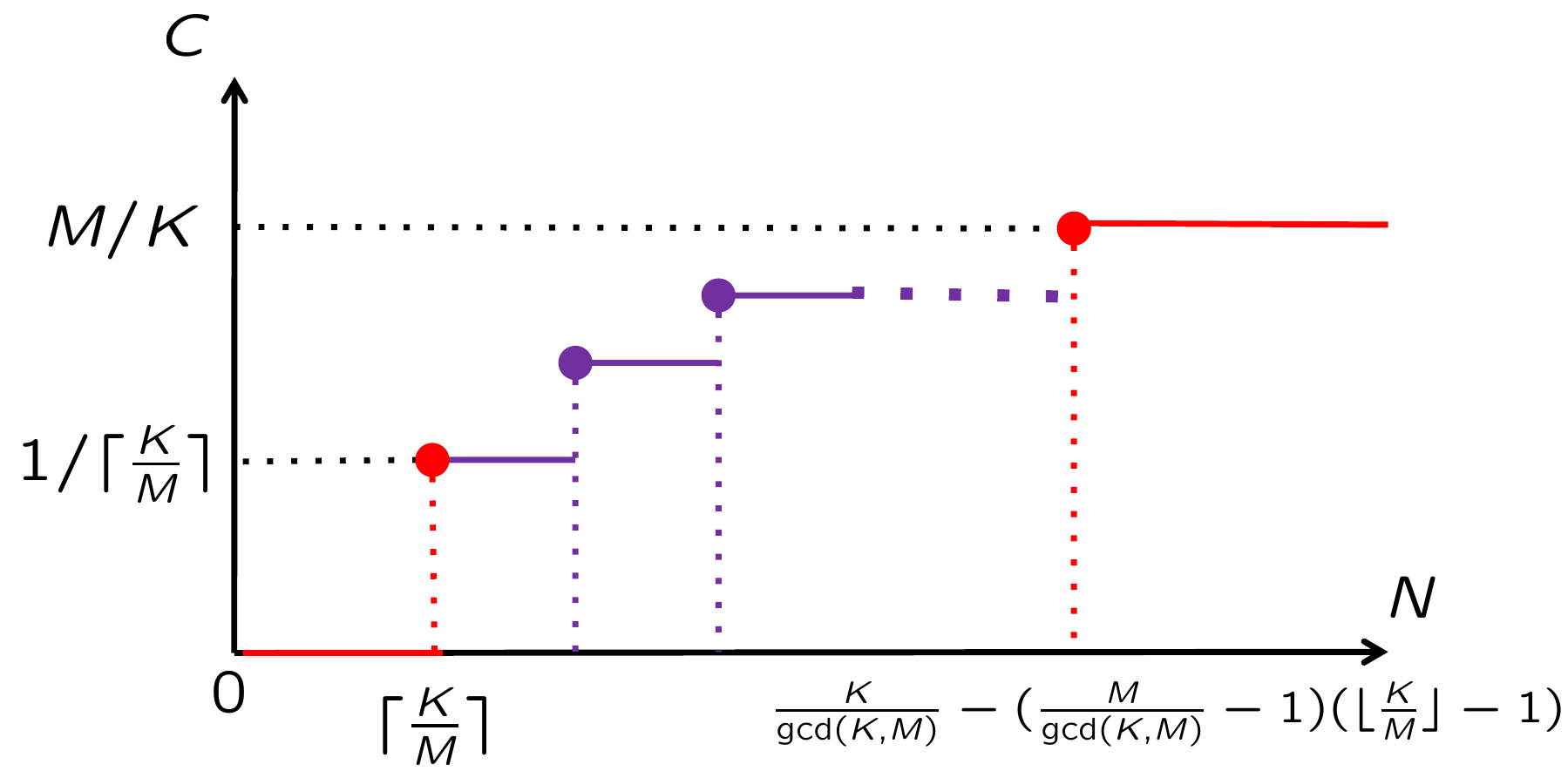}
\caption{\small The PID capacity. When $N \leq \lceil \frac{K}{M} \rceil$ and $N \geq \frac{K}{\gcd(K,M)} - (\frac{M}{\gcd(K,M)}-1)(\lfloor \frac{K}{M} \rfloor - 1)$, the capacity is fully characterized (colored in red) and otherwise, the capacity is open in general (colored in purple).}
\label{fig:sum}
\end{center}
\end{figure}

The results obtained so far are summarized in Figure \ref{fig:sum}. Beyond the intermediate regime $\lceil \frac{K}{M} \rceil < N < \frac{K}{\gcd(K,M)} - (\frac{M}{\gcd(K,M)}-1)(\lfloor \frac{K}{M} \rfloor - 1)$, we have characterized the PID capacity. %where the bounds in Theorem \ref{thm:approx} are not known to be tight. 
The range of $N$ in this regime is at most $\frac{2M}{\gcd(K,M)} - 1$, i.e.,
{\footnotesize
\begin{eqnarray}
&& \frac{K}{\gcd(K,M)} - (\frac{M}{\gcd(K,M)}-1)(\lfloor \frac{K}{M} \rfloor - 1) - \lceil \frac{K}{M} \rceil \notag \\
&=& \frac{M}{\gcd(K,M)} \frac{K}{M} - \frac{M}{\gcd(K,M)}(\lfloor \frac{K}{M} \rfloor -1) + \lfloor \frac{K}{M} \rfloor - \lceil \frac{K}{M} \rceil - 1\\
&=& \frac{M}{\gcd(K,M)}(\frac{K}{M} - \lfloor \frac{K}{M} \rfloor +1) + \lfloor \frac{K}{M} \rfloor - \lceil \frac{K}{M} \rceil - 1 < \frac{2M}{\gcd(K,M)} - 1
\end{eqnarray}}
Finally, we consider this intermediate regime and present an improved achievable scheme, in the following theorem.

\vspace{-0.3in}

\begin{theorem}\label{thm:ach}
For the private information delivery problem with $K$ messages, $N$ servers, and $M$ messages per server, when $\frac{K}{M} \notin \mathbb{Z}$ and $\lceil \frac{K}{M} \rceil < N < \frac{K}{\gcd(K,M)} - (\frac{M}{\gcd(K,M)}-1)(\lfloor \frac{K}{M} \rfloor - 1)$, the capacity satisfies
\begin{eqnarray}
C \geq \frac{l}{N + (l-1)(\lfloor \frac{K}{M} \rfloor - 1)},~\mbox{where}~ 
l = \lfloor \frac{(N - \lfloor \frac{K}{M} \rfloor + 1)M}{K - (\lfloor \frac{K}{M} \rfloor - 1) M} \rfloor.
\end{eqnarray}
\end{theorem}

\begin{figure}[h]
\begin{center}
\includegraphics[width=5 in]{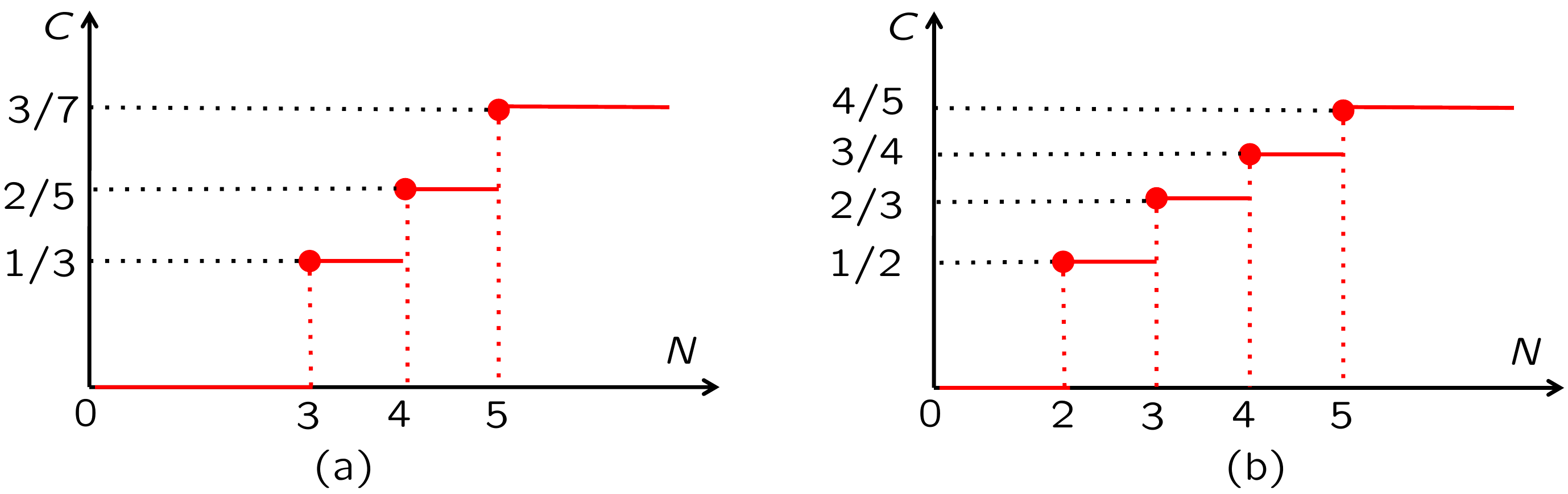}
\caption{\small The PID capacity, (a) when $M = 3, K = 7$, and (b), when $M = 4, K = 5$.}
\label{fig:c34}
\end{center}
\end{figure}

The proof of Theorem \ref{thm:ach} is presented in Section \ref{sec:ach}. To illustrate Theorem \ref{thm:ach}, we give two examples. 
\begin{example}
Suppose $M = 3, K = 7$. The only $N$ value that is covered in Theorem \ref{thm:ach} is $N = 4$. The achievable rate in Theorem \ref{thm:ach} is $2/5$. It turns out that this achievable rate is also optimal (proof deferred to Section \ref{sec:ex1}). Therefore, we have characterized the capacity of PID for all possible values of $N$ when $M=3, K=7$. This result is plotted in Figure \ref{fig:c34}(a).
\end{example}
%\begin{figure}[h]
%\begin{center}
%\includegraphics[width=2.5 in]{c3}
%\caption{\small The AID capacity when $M = 3, K = 7$.}
%\label{fig:c3}
%\end{center}
%\end{figure}

\begin{example}
Suppose $M = 4, K = 5$. The only $N$ values that are covered in Theorem \ref{thm:ach} are $N = 3, 4$. The achievable rate in Theorem \ref{thm:ach} is $2/3$ (when $N=3$), and $3/4$ (when $N=4$). It turns out that the achievable rates are also optimal (proof deferred to Section \ref{sec:ex2}). Therefore, we have characterized the capacity of PID for all possible values of $N$ when $M=4, K=5$. This result is plotted in Figure \ref{fig:c34}(b).
\end{example}

{\it Remark: The achievable rate in Theorem \ref{thm:ach} may not be monotonically increasing in $N$. So for a given $N$, if we want to find the highest achievable rate, we may search over all $N' \in [\lceil \frac{K}{M} \rceil + 1 : N]$.}

{\it Remark: The achievable scheme in Theorem \ref{thm:ach} includes those in Theorem \ref{thm:approx} and Theorem \ref{thm:tight} as special cases. That is, if we set $N = \lceil \frac{K}{M} \rceil$, the rate achieved in Theorem \ref{thm:ach} is $R = 1/\lceil \frac{K}{M} \rceil$ (same as that in Theorem \ref{thm:approx}), and if we set $N = \frac{K}{\gcd(K,M)} - (\frac{M}{\gcd(K,M)}-1)(\lfloor \frac{K}{M} \rfloor - 1)$, the rate achieved in Theorem \ref{thm:ach} is $R = M/K$ (same as that in Theorem \ref{thm:tight}).}

\section{Proof of Theorem \ref{thm:approx}}\label{sec:approx}
\subsection{Converse: $R \leq M/K$}
%\section{Theorem \ref{thm:main}: Converse} \label{sec:con}
Let us start with two useful lemmas. The first lemma states that if a message is available at a set of servers, then the size of the answers from these servers must be no less than the message size.
\begin{lemma}\label{lemma:basic}
Consider any storage strategy where $W_k$ is only available at servers in the set $\mathcal{N}_k = \{n_{k_1}, n_{k_2}, \cdots,n_{k_i}\}$, i.e., $W_k \in S_{j}, \forall  j \in \mathcal{N}_k$, and $W_k \notin S_l, \forall l \notin \mathcal{N}_k$. We have
\begin{eqnarray}
D_{\mathcal{N}_k^\Sigma} \define D_{n_{k_1}} + D_{n_{k_2}} + \cdots + D_{n_{k_i}} \geq L
\end{eqnarray}
\end{lemma}
{\it Proof:}
\begin{eqnarray}
L &\overset{(\ref{h1})}{=}& H(W_k) \\
&\overset{(\ref{corr})}{=}& I(W_k; A_1^{[k]}, A_2^{[k]}, \cdots, A_N^{[k]}) \\
&\overset{}{=}& I(W_k; A_{[1:N]/\mathcal{N}_k}^{[k]}) + I(W_k; A_{\mathcal{N}_k}^{[k]}|A_{[1:N]/\mathcal{N}_k}^{[k]})\\
&\overset{}{\leq}& I(W_k; A_{[1:N]/\mathcal{N}_k}^{[k]}, W_{[1:K]/k}, Z) + H(A_{\mathcal{N}_k}^{[k]}) \\
&\overset{(\ref{zk_ind})(\ref{privacy})}{\leq}& I(W_k; A_{[1:N]/\mathcal{N}_k}^{[k]} | W_{[1:K]/k}, Z) + D_{\mathcal{N}_k^\Sigma} \\
&\overset{(\ref{store})}{=}& I(W_k; A_{[1:N]/\mathcal{N}_k}^{[k]} | W_{[1:K]/k}, Z, S_{[1:N]/\mathcal{N}_k}^{[k]}) + D_{\mathcal{N}_k^\Sigma} \label{eq:s1}\\
&\overset{(\ref{det})}{=}& D_{\mathcal{N}_k^\Sigma}
\end{eqnarray}
where (\ref{eq:s1}) follows from the constraint that $W_k$ is not available at Server $l, \forall l \in [1:N]/\mathcal{N}_k$ so that $S_{[1:N]/\mathcal{N}_k}^{[k]} \subset W_{[1:K]/k}$. 
\hfill\QED

The second lemma states that having multiple servers storing the same set of messages does not help to reduce the private delivery rate.
\begin{lemma}\label{lemma:no_repl}
Consider any storage strategy $S_1, S_2, \cdots, S_N$ with $N' \leq N$ distinct $S_i$ storage variables. Without loss of generality, assume $S_i \neq S_j, \forall i\neq j, i, j \in [1:N']$. Then any rate $R$ that is achievable with $N$ servers and the storage strategy $S_1, S_2, \cdots, S_N$ is also achievable with $N'$ servers and the storage strategy $S_1, S_2, \cdots, S_{N'}$.
\end{lemma}

{\it Proof:} Suppose we are given a PID scheme (described by $NK$ answers $A_n^{[k]}, n \in [1:N], k \in [1:K]$) that operates over $N$ servers with the storage strategy $S_1, S_2, \cdots, S_N$, where $S_1, \cdots, S_{N'}$ are distinct. Denote the set of server indices for which the storage variables are equal to $S_i, i \in [1:N']$ by ${\mathcal{M}_i}$, i.e., if $j \in \mathcal{M}_i$, then $S_j = S_i$. Then we have $N'$ disjoint $\mathcal{M}_i$ sets that form a partition of the $N$ servers, i.e., $\mathcal{M}_1 \cup \mathcal{M}_2 \cup \cdots \mathcal{M}_{N'} = [1:N]$, and $\mathcal{M}_{i_1} \cap \mathcal{M}_{i_2} = \emptyset, \forall i_1 \neq i_2, i_1, i_2 \in [1:N']$.

Next we will construct a PID scheme that operates over $N'$ servers with the storage strategy $S_1, S_2, \cdots, S_{N'}$ and achieves the same rate as the $N$-server scheme above. We will use notations with a tilde symbol to describe the $N'$-server scheme. The common random variable remains the same, $\widetilde{Z} = Z$. The answer from Server $i$ to deliver $\widetilde{W}_k$ is denoted by $\widetilde{A}_i^{[k]}$. We set
\begin{eqnarray}
\widetilde{A}_i^{[k]} =  {A}_{\overrightarrow{\mathcal{M}_i}}^{[k]}, ~\forall n \in [1:N'], k \in [1:K] \label{eq:l1}
\end{eqnarray}
where $\overrightarrow{\mathcal{M}_i}$ is a vector that is in increasing order of the elements in the set ${\mathcal{M}_i}$. Note that the storage variable of all servers in the set $\mathcal{M}_i$ is the same as that of Server $i$, so that we may set the answers as above (refer to (\ref{det})). After collecting all $N'$ answers, we have
\begin{eqnarray}
(\widetilde{A}_1^{[k]}, \cdots, \widetilde{A}_{N'}^{[k]}) = ~\mbox{A permutation of}~(A_1^{[k]}, \cdots, A_N^{[k]}) \label{eq:l2}
\end{eqnarray}
so that we may use the decoding mapping (the order of the arguments in the mapping is correspondingly permuted) from the $N$-server scheme to decode $\widetilde{W}_k$. From (\ref{eq:l1}) and (\ref{eq:l2}), it is easy to see that the privacy constraint inherits and the same rate is preserved. The proof is therefore complete.
\hfill\QED

%\subsection{The proof of $R \leq M/K$ when $N \geq \frac{K}{\gcd(K,M)}$}
We are now ready to show that $R \leq M/K$. %when $N \geq \frac{K}{\gcd(K,M)}$.
From Lemma \ref{lemma:no_repl}, we may assume without loss of generality that the storage variables $S_n, n \in [1:N]$ are distinct. Note that $S_n$ is comprised of $M$ out of $K$ messages, so we have at most $\binom{K}{M}$ distinct $S_n$ variables. In other words, we may assume $N = \binom{K}{M}$ (note that having more servers can not hurt). Further, suppose the sets of stored messages $\mathcal{S}_1, \cdots, \mathcal{S}_{N}$ are ordered lexicographically. Consider any message $W_k, k \in [1:K]$, and $W_k$ is available at $\binom{K-1}{M-1}$ servers and this set of servers is denoted by $\mathcal{N}_k$, where $|\mathcal{N}_k| = \binom{K-1}{M-1}$. From Lemma \ref{lemma:basic}, we have
\begin{eqnarray}
L \leq D_{\mathcal{N}_k^\Sigma}  \label{eq:c1}
\end{eqnarray}
Adding (\ref{eq:c1}) for all $k \in [1:K]$, we have
\begin{eqnarray}
KL &\leq& \sum_{k=1}^K D_{\mathcal{N}_k^\Sigma} = M \sum_{n=1}^N D_n
\end{eqnarray}
where the last step follows from symmetry and any $D_n, n \in [1:N]$ appears $M$ times (Server $n$ contains $M$ messages and $D_n$ appears once for each message). Rearranging terms gives us the rate bound and completes the proof:
\begin{eqnarray}
R \overset{}{=} \frac{L}{\sum_{n=1}^N D_n} \leq M/K.
\end{eqnarray}

\subsection{Achievability: $R \geq 1/{\lceil \frac{K}{M} \rceil}$}%when $\frac{K}{\gcd(K,M)} > N \geq \lceil \frac{K}{M} \rceil$}
%When $\frac{K}{\gcd(K,M)} > N \geq \lceil \frac{K}{M} \rceil$, we show that the rate of $R = 1/{\lceil \frac{K}{M} \rceil}$ is achievable. As the rate does not depend on $N$, we only need to 
We provide a scheme with $N = \lceil \frac{K}{M} \rceil$ servers. 
Suppose each message is comprised of $L = 1$ symbol from $\mathbb{F}_2$ (in fact, any field will work).
The common random variable $Z$ consists of $N-1$ i.i.d. symbols, each from the same field $\mathbb{F}_2$. We denote $Z = (z_1, \cdots, z_{N-1})$.

The storage design is trivial, where the messages are stored sequentially over the servers.
\begin{eqnarray}
S_1 &=& \{{W}_1, {W}_2, \cdots, {W}_M\} \\%\overline{W}_{\frac{M}{\gcd(K,M)}}\} \\
S_2 &=& \{{W}_{M+1}, {W}_{M+2}, \cdots, {W}_{2M}\} \\%\overline{W}_{\frac{M}{\gcd(K,M)}+1}\} \\
 &\vdots& \\
S_{N} &=& \{{W}_{(N-1)M + 1}, \cdots, {W}_{K}\}
%S_{\frac{K}{\gcd(K,M)}} = \{\overline{W}_{\frac{K}{\gcd(K,M)}}, \overline{W}_1, \cdots, \overline{W}_{\frac{M}{\gcd(K,M)}-1}\} 
\end{eqnarray}

Suppose $W_k, k \in [1:K]$ is desired. The delivery scheme is linear, and each answer has $D_i = 1, \forall i \in [1:N]$ symbol. Then the rate achieved is $R = L/\sum_{i} D_i = 1/N = 1/{\lceil \frac{K}{M} \rceil}$, as desired.
The answers are shown below.
\begin{eqnarray}
A_i^{[k]} &=& z_i + 1(k \in [(i-1)M+1: iM]) W_k, ~~i \in [1:N-1], \\
A_N^{[k]} &=& -z_1 - \cdots - z_{N-1} + 1(k \in [(N-1)M+1: K]) W_k
\end{eqnarray}
where $1(x)$ denotes the indicator function that is equal to 1 if $x$ is true and 0 otherwise. Note that the answering symbol from Server $i$ contains $W_k$ only if $W_k$ is available at Server $i$.

To decode the desired message symbols, we add the $N$ answering symbols.
\begin{eqnarray}
W_k = A_1^{[k]} + A_2^{[k]} + \cdots + A_N^{[k]}
\end{eqnarray}
where all common randomness cancels and the desired message retains as it only appears once (in the answer from the server where it is stored). Note that the same decoding mapping is used for all $k$.

We next show that the privacy constraint (\ref{privacy}) is satisfied. To this end, note that regardless of the vale of the desired message index $k$, the answers are uniformly random, i.e., 
\begin{eqnarray}
H({A}_1^{[k]}, \cdots, A_N^{[k]}) = N.
\end{eqnarray}
Therefore, the scheme is both correct and private.

Finally, we note that $N-1$ randomness symbols are used to send $L = 1$ message symbol. The randomness size is then $\eta = H(Z)/L = N-1 = 1/R - 1$.

\section{Proof of Theorem \ref{thm:tight}}\label{sec:tight}
\subsection{$N = \frac{K}{\gcd(K,M)} - (\frac{M}{\gcd(K,M)}-1)(\lfloor \frac{K}{M} \rfloor - 1)$: Achievability of $R = M/K$}
%When $N \geq \frac{K}{\gcd(K,M)}$, we show that the rate of $R = M/K$ is achievable. 
%As the rate does not depend on $N$, we only need to provide a scheme with $N = \frac{K}{\gcd(K,M)}$ servers. 
\subsubsection{Example with $K = 8, M = 3$}
To illustrate the main idea in a simpler setting, we first consider an example with $K = 8, M = 3$ so that $N = 8 - (3-1)(2-1)= 6$ and we show that $R = M/K = 3/8$ is achievable.

Suppose the message size $L = 3$ symbols, and each symbol is from $\mathbb{F}_p$, where $p \geq 8$. Then $W_k, k \in [1:8]$ is a $3 \times 1$ vector. The common random variable consists of 5 i.i.d. symbols from the same field $\mathbb{F}_p$, i.e., $Z \in \mathbb{F}_p^{5\times 1}$. The storage is designed as follows, where the first $5$ servers store $M=3$ messages out of $W_1, W_2, W_3, W_4, W_5$ in a cyclic manner and the last server stores the remaining $3$ messages $W_6, W_7, W_8$.
\begin{eqnarray}
S_1 &=& \{W_{1}, W_2, W_3\} \\
S_2 &=& \{W_{2}, W_3, W_4\} \\
S_3 &=& \{W_{3}, W_4, W_5\} \\
S_4 &=& \{W_{4}, W_5, W_1\} \\
S_5 &=& \{W_{5}, W_1, W_2\} \\
S_6 &=& \{W_{6}, W_7, W_8\} 
\end{eqnarray}

Let us start with the case where $W_1$ is desired. The delivery scheme is linear, and the first 5 answer has $D_i = 1, \forall i \in [1:5]$ symbol each while the last answer has $D_6 = 3$ symbols. Then the rate achieved is $R = L/\sum_{i} D_i = 3/8$, as desired.
The collection of the answers is shown below. Define $W_1 = (a_1, a_2, a_3), Z = (z_1, z_2, z_3, z_4, z_5)$.
\begin{eqnarray}
{\bf A}^{[1]} \triangleq 
\left[\begin{array}{c}
A_1^{[1]} \\
A_2^{[1]} \\
A_3^{[1]} \\
A_4^{[1]} \\
A_5^{[1]} \\
A_6^{[1]}
\end{array}\right] = 
\left[\begin{array}{cc}
{\bf f}_1^{[1]} & {\bf h}_1\\
{\bf 0}_{1 \times 3} & {\bf h}_2\\
{\bf 0}_{1 \times 3} & {\bf h}_3\\
{\bf f}_4^{[1]} & {\bf h}_4\\
{\bf f}_5^{[1]} & {\bf h}_5 \\
{\bf 0}_{3 \times 3} & {\bf H}_{6}
\end{array}\right]
\left[\begin{array}{c}
a_1 \\
a_2 \\
a_3 \\
z_1 \\
z_2 \\
z_3 \\
z_4 \\
z_5
\end{array}\right] \label{eq:a1}
%A_1^{[1]} &=& {\bf f}_1^{[1]} W_1 + {\bf h}_1 Z \\
%A_2^{[1]} &=& {\bf f}_1^{[1]} W_1 + {\bf h}_2 Z \\
%A_3^{[1]} &=& {\bf f}_1^{[1]} W_1 + {\bf h}_3 Z \\
%A_4^{[1]} &=& {\bf f}_1^{[1]} W_1 + {\bf h}_4 Z \\
%A_5^{[1]} &=& {\bf f}_1^{[1]} W_1 + {\bf h}_5 Z \\
\end{eqnarray}
where in answer $A_i^{[1]}, i \in [1:5]$ from Server $i$, ${\bf f}_i^{[1]}$ is a $1 \times 3$ precoding vector for the message symbols $W_1 \in \mathbb{F}_p^{3\times 1}$ and ${\bf h}_i$ is a $1 \times 5$ precoding vector for the common randomness symbols $Z \in \mathbb{F}_p^{5\times1}$. In answer $A_6^{[1]}$, the $3\times 3$ precoding matrix ${\bf F}_6^{[1]}$ for $W_1$ is set as the zero matrix and  ${\bf H}_6$ is the $3 \times 5$ precoding matrix for $Z$.
Note that as $W_1$ is not stored at Servers $2, 3, 6$, ${\bf f}_2^{[1]}, {\bf f}_3^{[1]}, {\bf F}_6^{[1]}$ must be zero. It turns out that in our scheme, the precoding vectors for the common randomness do not depend on the desired message index. Define
\begin{eqnarray}
{\bf F}^{[1]}_{8\times 3} \triangleq
\left[\begin{array}{cc}
{\bf f}_1^{[1]}\\
{\bf 0}_{3\times 1} \\
{\bf 0}_{3\times 1} \\
{\bf f}_4^{[1]}  \\
{\bf f}_5^{[1]}  \\
{\bf 0}_{3\times 3}
\end{array}\right], ~~
{\bf H}_{8\times 5} \triangleq
\left[\begin{array}{cc}
{\bf h}_1\\
{\bf h}_2 \\
{\bf h}_3 \\
{\bf h}_4 \\
{\bf h}_5 \\
{\bf H}_6
\end{array}\right]
\end{eqnarray}
and (\ref{eq:a1}) may be re-written as
\begin{eqnarray}
{\bf A}^{[1]} = [{\bf F}^{[1]}~~ {\bf H}] 
\left[\begin{array}{c}
W_1\\
Z
\end{array}
\right] = {\bf F}^{[1]} W_1 + {\bf H} Z.
\end{eqnarray}
To decode the 3 desired message symbols from the 8 answering symbols, we apply a $3 \times 8$ linear filtering matrix ${\bf G}_{3 \times 8}$ to ${\bf A}^{[1]}$. We have
\begin{eqnarray}
W_1 = 
\left[\begin{array}{c}
a_1 \\
a_2 \\
a_3
\end{array}
\right] = 
{\bf G} {\bf A}^{[1]} = {\bf G} {\bf F}^{[1]} W_1 + {\bf G} {\bf H} Z, \label{eq:d1}
\end{eqnarray}
and to satisfy (\ref{eq:d1}), we set
\begin{eqnarray}
{\bf G} {\bf F}^{[1]} &=& {\bf I}_{3} \Rightarrow {\bf G}_{[:, (1,4,5)]} {\bf F}^{[1]}_{[(1,4,5), :]} = {\bf I}_3, \label{eq:con1}\\
{\bf G} {\bf H} &=& {\bf 0}_{3 \times 5} \label{eq:con2}. 
\end{eqnarray}
Note that ${\bf G}_{[:, (1,4,5)]}, {\bf F}^{[1]}_{[(1,4,5), :]}$ are both square matrices.

The situation where $W_k, k \in [2:8]$ is desired is similar. The answers are
\begin{eqnarray}
{\bf A}^{[k]} = {\bf F}^{[k]} W_k + {\bf H} Z
\end{eqnarray}
and the decoding constraints are (the answers are projected onto ${\bf G}$ to decode the desired message)
\begin{eqnarray}
{\bf G}_{[:, (1,2,5)]} {\bf F}^{[2]}_{[(1,2,5), :]} &=& {\bf I}_3, \label{eq:con3}\\
{\bf G}_{[:, (1,2,3)]} {\bf F}^{[3]}_{[(1,2,3), :]} &=& {\bf I}_3, \\
{\bf G}_{[:, (2,3,4)]} {\bf F}^{[4]}_{[(2,3,4), :]} &=& {\bf I}_3, \\
{\bf G}_{[:, (3,4,5)]} {\bf F}^{[5]}_{[(3,4,5), :]} &=& {\bf I}_3, \\
{\bf G}_{[:, (6,7,8)]} {\bf F}^{[j]}_{[(6,7,8), :]} &=& {\bf I}_3, j \in [6:8]
%\\
%{\bf G}_{[:, (6,7,8)]} {\bf F}^{[7]}_{[(6,7,8), :]} &=& {\bf I}_3, \\
%{\bf G}_{[:, (6,7,8)]} {\bf F}^{[8]}_{[(6,7,8), :]} &=& {\bf I}_3, 
\label{eq:con5}\\
{\bf G} {\bf H} &=& {\bf 0}_{3 \times 5}. \label{eq:con4}
\end{eqnarray}
Note that the same decoding mapping ${\bf G}$ must be used for each desired message. So the delivery design reduces to find a realization of the matrices ${\bf G}, {\bf F}^{[1]}, {\bf F}^{[2]}, \cdots, {\bf F}^{[8]}, {\bf H}$ such that (\ref{eq:con1}), (\ref{eq:con2}), (\ref{eq:con3}) - (\ref{eq:con4}) are satisfied.

These matrices are chosen as follows. We first set
\begin{eqnarray}
{\bf G} = \left[
\begin{array}{cc}
{\bf I}_3
&
{\bf V}_{3 \times 5}
\end{array}
\right]
\end{eqnarray}
where 
\begin{eqnarray}
{\bf V}_{3\times 5} = 
\left[ \begin{array}{ccccc}
\frac{1}{\alpha_1-\beta_1} & \frac{1}{\alpha_1-\beta_2} & \frac{1}{\alpha_1-\beta_3} & \frac{1}{\alpha_1-\beta_4} & \frac{1}{\alpha_1-\beta_5}\\
\frac{1}{\alpha_2-\beta_1} & \frac{1}{\alpha_2-\beta_2} & \frac{1}{\alpha_2-\beta_3} & \frac{1}{\alpha_2-\beta_4} & \frac{1}{\alpha_2-\beta_5}\\
\frac{1}{\alpha_3-\beta_1} & \frac{1}{\alpha_3-\beta_2} & \frac{1}{\alpha_3-\beta_3} & \frac{1}{\alpha_3-\beta_4} & \frac{1}{\alpha_3-\beta_5}
\end{array}
\right], \alpha_i, \beta_j, i\in[1:5], j \in[1:3]~\mbox{are all distinct.}%\alpha_{j_1} \notin \{0,1\}, \alpha_{j_1} \neq \alpha_{j_2}, \forall j_1, j_2 \in [1:5], j_1 \neq j_2.
\end{eqnarray}
It is guaranteed that we can find such $\alpha_{i}, \beta_j$ because the field size $p\geq8$. In other words, ${\bf V}$ is a Cauchy matrix where each square sub-matrix is invertible.
%We first choose each element of ${\bf G}$ i.i.d. and uniformly from a sufficiently large field. 
Then ${\bf H}$ is solved from (\ref{eq:con2}), as the right null space of ${\bf G}$. 
\begin{eqnarray}
{\bf H} = \left[
\begin{array}{cc}
{\bf V}_{3\times 5}\\
-{\bf I}_5
\end{array}
\right].
\end{eqnarray}
Next, the submatrices of ${\bf F}^{[k]}, k \in [1:8]$ are set as the inverse matrices of corresponding submatrices of ${\bf G}$, from (\ref{eq:con1}), (\ref{eq:con3}) - (\ref{eq:con5}). Note that it is easy to see the corresponding submatrices of ${\bf G}$ have full rank such that their inverse matrices exist. Then ${\bf F}^{[k]}$ are fully determined as the rows that have not appeared are zero vectors, due to the storage constraint.
Now all correctness constraints are satisfied. We are left to show that the privacy constraint (\ref{privacy}) is satisfied. To this end, we show that regardless of the vale of the desired message index $k$, the answers are uniformly random, which translates to that the following matrices have full rank.
\begin{eqnarray}
\mbox{(Equivalent Privacy Condition):} ~~~ {\bf B}^{[k]}_{8 \times 8} \triangleq [{\bf F}^{[k]} ~~{\bf H}], ~\forall k \in [1:8] ~\mbox{have full rank}.
\end{eqnarray}
As each ${\bf F}^{[k]}$ contains 5 zero rows, it suffices to show that any 5 rows of ${\bf H}$ are linearly independent (holds trivially by the determinant formula of Cauchy matrices). A more detailed proof will be presented in the general proof.

The construction of the matrices is not unique. In fact, it is not hard to show that if we choose each element of ${\bf G}$ i.i.d. and uniformly from a sufficiently large field and follow the above procedure to determine ${\bf F}^{[k]}$ and ${\bf H}$, then the solution will work with a high probability.

Finally, we note that 5 randomness symbols are used to send 3 message symbols. The randomness size is then $\eta = H(Z)/L = 5/3 = 1/R - 1$.

\subsubsection{General proof with arbitrary $K, M$}\label{sec:general}
We show that for $K$ messages, $M$ messages per server, and $N = \frac{K}{\gcd(K,M)} - (\frac{M}{\gcd(K,M)}-1)(\lfloor \frac{K}{M} \rfloor - 1)$ servers, the rate $R = M/K$ is achievable.

We treat every $\gcd(K,M)$ messages as a block so that we have $\overline{K} \triangleq \frac{K}{\gcd(K,M)}$ message blocks. Define
\begin{eqnarray}
\overline{W}_b = \{W_{(b-1)\gcd(K,M) +1}, W_{(b-1)\gcd(K,M) +2}, \cdots, W_{b\gcd(K,M)}\}, b \in [1: \overline{K}]. %\frac{K}{\gcd(K,M)}]
\end{eqnarray}
Each server now is able to store $\overline{M} \triangleq \frac{M}{\gcd(K,M)}$ message blocks. 

Suppose the message size $L = \overline{M}$ symbols. Each symbol is from $\mathbb{F}_p$, where $p\geq{\color{black}\overline{K}}$.
The common random variable $Z$ consists of ${\color{black}\overline{K}-L}$ i.i.d. symbols, each from the same field $\mathbb{F}_p$. 

We divide the $N$ servers into 2 sets. The first set is made up of the first $N_1$ servers and the second set is made up of the last $N_2$ servers, where
\begin{eqnarray}
N_2 &=& \lfloor \frac{{K}}{{M}} \rfloor - 1 \\
N_1 &=& N - N_2
\end{eqnarray} 
The message blocks also are divided into 2 sets, where the first set is comprised of the first $N_1$ message blocks and the second set is comprised of the remaining $\overline{K} - N_1$ message blocks.

The storage is designed as follows. In the first server set, each server stores $L$ ($=\overline{M}$) message blocks out of the first message set in a cyclic manner. In the second server set, each server stores $L$ distinct message blocks from the second message set sequentially.
\begin{eqnarray}
S_1 &=& \{\overline{W}_1, \overline{W}_2, \cdots, \overline{W}_L\} \\%\overline{W}_{\frac{M}{\gcd(K,M)}}\} \\
S_2 &=& \{\overline{W}_2, \overline{W}_3, \cdots, \overline{W}_{L+1}\} \\%\overline{W}_{\frac{M}{\gcd(K,M)}+1}\} \\
 &\vdots& \\
S_{N_1} &=& \{\overline{W}_{N_1}, \overline{W}_1, \cdots, \overline{W}_{L-1}\} \\
S_{N_1 + 1} &=& \{\overline{W}_{N_1+1}, \overline{W}_{N_1+2}, \cdots, \overline{W}_{N_1 + L}\} \\
 &\vdots& \\
S_{N} &=&  \{\overline{W}_{N_1+(N_2 - 1)L+1}, \cdots, \overline{W}_{N_1 + N_2 L}\}
%S_{\frac{K}{\gcd(K,M)}} = \{\overline{W}_{\frac{K}{\gcd(K,M)}}, \overline{W}_1, \cdots, \overline{W}_{\frac{M}{\gcd(K,M)}-1}\} 
\end{eqnarray}
To see that all messages are stored, we show that the last message block $\overline{W}_{N_1 + N_2 L}$ is indeed $\overline{W}_{\overline{K}}$,
\begin{eqnarray}
N_1 + N_2 L &=& (N-N_2) + N_2 \overline{M}\\
&=& N + (\lfloor \frac{{K}}{{M}} \rfloor - 1) (\overline{M} -1) ~~~~\mbox{(Using the definition of $N_2$)}\\
&=& \overline{K}.~~~~~~~~~~~~~~~~~~~~~~~~~~~~~~~~\mbox{(Using the definition of $N$)}  \label{eq:kbar}
\end{eqnarray}

Suppose $W_k, k \in [1:K]$ is desired. The delivery scheme is linear, where each answer from the first server set has $D_i = 1, \forall i \in [1:N_1]$ symbol, and each answer from the second server set has $D_i = L, \forall i \in [N_1 + 1: N]$ symbols. Then the rate achieved is 
\begin{eqnarray}
R = \frac{L}{\sum_{i} D_i} = \frac{L}{N_1 + L N_2} \overset{(\ref{eq:kbar})}{=} \overline{M}/\overline{K} = M/K
\end{eqnarray}
and it matches the desired rate expression.
The answers are shown below.
\begin{eqnarray}
{A}_i^{[k]} &=& {\bf F}_i^{[k]} W_k + {\bf H}_i Z
\end{eqnarray}
where if $i \in [1:N_1]$, ${\bf F}_i^{[k]}$ has dimension $1 \times L$, ${\bf H}_i$ has dimension $1 \times (\overline{K} - L)$, and otherwise if $i \in [N_1 + 1:N]$, ${\bf F}_i^{[k]}$ has dimension $L \times L$, ${\bf H}_i$ has dimension $L \times (\overline{K} - L)$. Define
\begin{eqnarray}
{\bf F}^{[k]}_{\overline{K} \times L} &=& [{\bf F}_1^{[k]}; {\bf F}_2^{[k]}; \cdots; {\bf F}_N^{[k]}], \\
{\bf H}_{\overline{K} \times (\overline{K}-L)} &=& [{\bf H}_1; {\bf H}_2; \cdots; {\bf H}_N]
\end{eqnarray}
and we have the collection of all answers,
\begin{eqnarray}
{\bf A}^{[k]} &=& [{\bf F}^{[k]}~~ {\bf H}] 
\left[\begin{array}{c}
W_k\\
Z
\end{array}
\right] = {\bf F}^{[k]} W_k + {\bf H} Z. \label{eq:ge1}
\end{eqnarray}
We next specify the availability set $\mathcal{N}_k$ of $W_k$, i.e., $W_k$ is only available at Server $n$ where $n \in \mathcal{N}_k$. Note that $W_k$ belongs to message block $\overline{W}_{\overline{k}}$, where $\overline{k} \triangleq \lceil \frac{k}{\gcd(K,M)} \rceil$.
\begin{eqnarray}
\mathcal{N}_k = \left\{ 
\begin{array}{cl}
[\overline{k} - L+1: \overline{k}]\mod N_1 & \mbox{if}~ \overline{k} \in [1: N_1], \\
\lceil{(\bar{k} - N_1)}/{\overline{M}}\rceil  & \mbox{else}~ \overline{k} \in [N_1 + 1: \overline{K}].\\
\end{array}
\right.
\end{eqnarray}
Due to the above storage constraints, we have the following corresponding constraints on the precoding matrices.
\begin{eqnarray}
\begin{array}{lc}
 \mbox{If}~\overline{k} \in [1: N_1], & {\bf F}^{[k]}_{n_1} = {\bf 0}_{1 \times L}, \forall n_1 \notin \mathcal{N}_k, n_1 \in [1:N_1], ~{\bf F}^{[k]}_{n_2} = {\bf 0}_{L \times L}, \forall n_2 \in [N_1 + 1: N] \\
 \mbox{else}~\overline{k} \in [N_1 + 1: \overline{K}], & {\bf F}^{[k]}_{n_1} = {\bf 0}_{1 \times L}, \forall n_1 \in [1:N_1], ~{\bf F}^{[k]}_{n_2} = {\bf 0}_{L \times L}, \forall n_2 \notin \mathcal{N}_k, n_2 \in [N_1 + 1: N]
\end{array}
\label{eq:gz1}
\end{eqnarray}

To decode the $L$ desired message symbols from the $\overline{K}$ answering symbols, we apply a linear filtering matrix ${\bf G}_{L \times \overline{K}}$ to ${\bf A}^{[k]}$. We have
\begin{eqnarray}
W_k = 
{\bf G} {\bf A}^{[k]} = {\bf G} {\bf F}^{[k]} W_k + {\bf G} {\bf H} Z, \label{eq:gd1}
\end{eqnarray}
and to satisfy (\ref{eq:gd1}), we set
\begin{eqnarray}
{\bf G} {\bf F}^{[k]} &=& {\bf I}_{L} \Rightarrow 
\begin{array}{ll} 
\mbox{If}~\overline{k} \in [1: N_1], & {\bf G}_{[:, \overrightarrow{\mathcal{N}}_k]} {\bf F}^{[k]}_{[\overrightarrow{\mathcal{N}}_k, :]} = {\bf I}_L, \\
 \mbox{else}~\overline{k} \in [N_1 + 1: \overline{K}], & {\bf G}_{[:, N_1+ ({\mathcal{N}}_k -N_1-1)L + 1: N_1 + ({\mathcal{N}}_k -N_1) L]} {\bf F}^{[k]}_{{\mathcal{N}}_k} = {\bf I}_L 
\end{array} 
\label{eq:gcon1} \\
{\bf G} {\bf H} &=& {\bf 0}_{L \times (\overline{K}-L)} \label{eq:gcon2}
\end{eqnarray}
where the vector $\overrightarrow{\mathcal{N}}_k$ is in increasing order of elements in the set $\mathcal{N}_k$ (the available set for message $W_k$). For example, suppose $M = 6, K = 20, k = 2$. Then $\gcd(M,K) = 2, N = 6, N_1 = 4, N_2 = 2, L = 3, \overline{k} = 1$, $\mathcal{N}_2 = \{3,4,1\}$, and $\overrightarrow{\mathcal{N}}_2 = (1,3,4)$.

We next find matrices ${\bf G}, {\bf F}^{[1]}, {\bf F}^{[2]}, \cdots, {\bf F}^{[K]}, {\bf H}$ such that (\ref{eq:gcon1}), (\ref{eq:gcon2}) are satisfied for all $k \in [1:K]$.
We first set
\begin{eqnarray}
{\bf G} = \left[
\begin{array}{ccccc}
{\bf I}_{L} & {\bf V}_{L \times (\overline{K}-L)}
\end{array}
\right]
\end{eqnarray}
where ${\bf V}_{L \times (\overline{K}-L)}$ is a Cauchy matrix such that the element in $i$-th row and $j$-th column is given by
\begin{eqnarray}
V_{ij} = \frac{1}{\alpha_i - \beta_j}
\end{eqnarray}
and $\alpha_i, \beta_j$ are distinct elements over $\mathbb{F}_p$ where $p \geq \overline{K}$.
%\begin{eqnarray}
%{\bf V}_{L \times (\overline{K}-L)} \triangleq 
%\left[\begin{array}{cccccc}
%1 & 1 & \cdots & 1 \\
%\alpha_1 & \alpha_2 & \cdots & \alpha_{\overline{K}-L} \\
%\vdots & \vdots & \vdots & \vdots \\
%\alpha_1^{L-1} & \alpha_2^{L-1} & \cdots & \alpha_{\overline{K}-L}^{L-1}
%\end{array}
%\right], ~\alpha_{j_1} \in \mathbb{F}_p/\{0, 1\}, \alpha_{j_1} \neq \alpha_{j_2}, \forall j_1, j_2 \in [1:\overline{K}-L], j_1 \neq j_2.
%\end{eqnarray}
Then ${\bf H}$ is solved from (\ref{eq:gcon2}) as the right null space of ${\bf G}$. The non-zero rows of ${\bf F}^{[k]}$ are solved from (\ref{eq:gcon1}), as the inverse of some sub-matrices of ${\bf G}_{}$. 
\begin{eqnarray}
&&{\bf H} = \left[
\begin{array}{cc}
{\bf V}_{L \times (\overline{K}-L)} \\
-{\bf I}_{\overline{K}-L}
\end{array}
\right], \\
&& \begin{array}{ll}
{\bf F}^{[k]}_{[\overrightarrow{\mathcal{N}}_k, :]} = {\bf G}_{[:, \overrightarrow{\mathcal{N}}_k]}^{-1}, & \mbox{if}~\overline{k} \in [1: N_1], \\
{\bf F}^{[k]}_{{\mathcal{N}}_k} = {\bf G}_{[:, N_1+ ({\mathcal{N}}_k -N_1 -1)L + 1: N_1 + ({\mathcal{N}}_k - N_1) L]}^{-1},  &  \mbox{else}~\overline{k} \in [N_1 + 1: \overline{K}].
\end{array}\label{eq:ninv}
\end{eqnarray}
Note that if $\overline{k} \in [1: N_1]$, $\overrightarrow{\mathcal{N}}_k$ consists of $L$ cyclicly consecutive elements in $[1:N_1]$ such that ${\bf G}_{[:, \overrightarrow{\mathcal{N}}_k]}$ is non-singular (its determinant is equal to the determinant of a square Cauchy matrix), and otherwise if $\overline{k} \in [N_1 + 1: \overline{K}]$, ${\bf G}_{[:, N_1+ ({\mathcal{N}}_k -N_1-1)L + 1: N_1 + ({\mathcal{N}}_k - N_1)L]}$ consists of $L$ consecutive columns from ${\bf G}$ and is non-singular as well.

Now all correctness constraints are satisfied. We are left to show that the privacy constraint (\ref{privacy}) is satisfied. To this end, we show that regardless of the vale of the desired message index $k$, the answers are uniformly random, i.e., 
\begin{eqnarray}
H({\bf A}^{[k]}) = \overline{K} = H(W_k, Z).
\end{eqnarray}
From (\ref{eq:ge1}), it is equivalent to show that
\begin{eqnarray}
\mbox{(Equivalent Privacy Condition):} ~~~ {\bf B}^{[k]}_{\overline{K} \times \overline{K}} = [{\bf F}^{[k]} ~~{\bf H}], ~\forall k \in [1:K] ~\mbox{have full rank}.
\end{eqnarray}

First, consider the case where $\overline{k} \in [1: N_1]$.
From (\ref{eq:gz1}), we know that $N-L$ cyclicly consecutive rows (where the row index does not belong to the set $\mathcal{N}_k$) of ${\bf F}^{[k]}$ are the zero vectors. It follows from the determinant formula of a $2\times 2$ block matrix with a zero sub-block that
\begin{eqnarray}
\det({\bf B}^{[k]}) = \det({\bf F}^{[k]}_{[\overrightarrow{\mathcal{N}}_k, :]}) \det({\bf H}_{[(1:\overline{K})/\overrightarrow{\mathcal{N}}_k,:]}) \label{eq:bl}
\end{eqnarray}
We have shown that ${\bf F}^{[k]}_{[\overrightarrow{\mathcal{N}}_k, :]}$ is non-singular. Further $|{\bf H}_{[(1:\overline{K})/\overrightarrow{\mathcal{N}}_k,:]}|$ is equal to the determinant of a square sub-matrix of a Cauchy matrix (and is another Cauchy matrix) so that ${\bf H}_{[(1:\overline{K})/\overrightarrow{\mathcal{N}}_k,:]}$ is non-singular as well. 

Second, consider the case where $\overline{k} \in [N_1 + 1: \overline{K}]$. The proof is similar to that above, where the non-zero part of the ${\bf F}^{[k]}$ component is a non-singular square matrix (refer to (\ref{eq:ninv})) and the corresponding sub-matrix of the ${\bf H}$ component in the determinant formula (refer to (\ref{eq:bl})) has a determinant that is given by a square sub-matrix of a Cauchy matrix (thus non-singular as well).
Therefore, ${\bf B}^{[k]}$ always have full rank and the scheme is private.

Finally, we note that $\overline{K}-L$ randomness symbols are used to send $L$ message symbols. The randomness size is then $\eta = H(Z)/L = (\overline{K}-L)/L = (K-M)/M = 1/R - 1$.

{\it Remark: An interesting observation of our scheme is that it is automatically secure, i.e., from the answers for $W_k$, the user learns absolutely no information about other messages. This indicates that the undelivered messages do not play a role in keeping privacy (the common randomness is responsible for privacy).}

\subsection{$N = \lceil \frac{K}{M} \rceil$: Optimality of $R = 1/\lceil \frac{K}{M} \rceil$}
%We assume that each server stores $M$ distinct messages, as storing more messages does not hurt
We first show that when $N = \lceil \frac{K}{M} \rceil$, each server must contain a message that appears only in that server (not available in any other servers). That is, we have 
\begin{enumerate}
\item[] (Property 1) For any $i \in [1:N]$, there exists a message $W_{k_i} \in S_i$, and $W_{k_i} \notin S_j, \forall j \neq i, j \in [1:N]$. %, and ${k_{i_1}} \neq k_{i_2}, \forall i_1, i_2 \in [1:N], i_1 \neq i_2$.
\end{enumerate}
We now prove that Property 1 holds. To set up the proof by contradiction, suppose there exists 1 server (say Server $n$) where every stored message appears at some other server. As each server stores $M$ messages, we know that the $M$ messages stored at Server $n$ are replicated at least twice. As a result, the total storage required at all $N$ servers is at least $K + M$. However, this is not possible because $K+M$ exceeds the total storage capability of the servers, $MN$.
\begin{eqnarray}
MN &=& M \times \lceil \frac{K}{M} \rceil \\
&<& M \times \left(\frac{K}{M} + 1\right) = K + M
\end{eqnarray}
So we have proved that Property 1 is satisfied. Consider these $N$ messages $W_{k_1}, \cdots, W_{k_N}$, where each of them is available at only 1 (distinct) server. Using Lemma \ref{lemma:basic} for $W_{k_i}$, we have
\begin{eqnarray}
\mathcal{N}_{k_i} = \{i\}: ~~D_{i} \geq L \label{eq:o1}
\end{eqnarray}
Adding (\ref{eq:o1}) for all $i \in [1:N]$, we have
\begin{eqnarray}
 D_1 + \cdots + D_N &\geq& NL \\
\Rightarrow R = \frac{L}{\sum_{n=1}^N D_n} &\leq& 1/N = 1/\lceil \frac{K}{M} \rceil
\end{eqnarray}
and the proof is complete.

\section{Proof of Theorem \ref{thm:ach}}\label{sec:ach}
The achievable scheme is similar to that presented in Section \ref{sec:general} (albeit with a different set of parameters). Here we present the code construction succinctly and only highlight the differences.

We show that for $K$ messages, $M$ messages per server, and $N$ servers, the following rate is achievable.
\begin{eqnarray}
R = \frac{l}{N + (l-1)(\lfloor \frac{K}{M} \rfloor - 1)},~\mbox{where}~l = \lfloor \frac{(N - \lfloor \frac{K}{M} \rfloor + 1)M}{K - (\lfloor \frac{K}{M} \rfloor - 1) M} \rfloor. \label{eq:ldef}
\end{eqnarray}

The $N$ servers and $K$ messages are similarly divided into 2 sets.
The first server set is made up of the first $N_1 = N - N_2$ servers and the second server set is made up of the last $N_2 = \lfloor \frac{{K}}{{M}} \rfloor - 1$ servers.
The first message set is comprised of the first $K_1 = K - K_2$ messages and the second message set is comprised of the last $K_2 = N_2 M$ messages.

Suppose the message size $L = l$ symbols. Define $D_\Sigma \triangleq N_1 + L N_2$. Each message symbol is from $\mathbb{F}_p$, where $p\geq{\color{black}D_{\Sigma}}$.
The common random variable $Z$ consists of ${\color{black}D_{\Sigma}-L}$ i.i.d. symbols, each from the same field $\mathbb{F}_p$. 

The storage is designed as follows. The first (second) message set is stored over the first (second) server set. Consider the first server set, where the $N_1$ servers can store $N_1 M$ messages. Note that there are $K_1$ messages in the first message set so that at least, each of these $K_1$ messages can be stored $l = \lfloor N_1M/K_1 \rfloor$ times (refer to (\ref{eq:ldef})). Imagine these $N_1 M$ locations as an $N_1 \times M$ table with $N_1$ rows and $M$ columns. Consider the $N_1M$ locations of the table in a greedily manner, first from the first row to the last row and then from the first column to the last column, and we throw the $K_1 l$ messages (from $W_1$ to $W_{K_1}$, each message replicated $l$ times) into the locations in the order specified. The desired property of this storage strategy is that each message $W_i, i \in [1:K-1]$ is available at $l$ cyclicly consecutive servers in the first server set. Denote the availability set of $W_k$ as $\mathcal{N}_k$. 
%We greedily consider the $K_1$ messages. Starting from $W_1$, we store it from the $(1,1)$-th location in the table to the $(l,1)$-th location.
In the second server set, each server stores $L$ distinct messages from the second message set sequentially.
%\begin{eqnarray}
%S_1 &=& \{{W}_1, \cdots \} \\%\overline{W}_{\frac{M}{\gcd(K,M)}}\} \\
%%S_2 &=& \{{W}_1, {W}_3, \cdots, {W}_{L+1}\} \\%\overline{W}_{\frac{M}{\gcd(K,M)}+1}\} \\
%&\vdots& \\
%S_l &=& \{{W}_1, \cdots\} \\%\overline{W}_{\frac{M}{\gcd(K,M)}+1}\} \\
%S_{l+1} &=& \{{W}_2, \cdots \} \\%\overline{W}_{\frac{M}{\gcd(K,M)}+1}\} \\
%S_{l+2} &=& \{{W}_2, \cdots \} \\%\overline{W}_{\frac{M}{\gcd(K,M)}+1}\} \\
% &\vdots& \\
%S_{N_1} &=& \{\cdots\} \\
%S_{N_1 + 1} &=& \{{W}_{K_1+1}, {W}_{K_1+2}, \cdots, {W}_{K_1 + M}\} \\
% &\vdots& \\
%S_{N} &=&  \{{W}_{K_1+(N_2 - 1)M+1}, \cdots, {W}_{K_1 + N_2 M}\}
%%S_{\frac{K}{\gcd(K,M)}} = \{\overline{W}_{\frac{K}{\gcd(K,M)}}, \overline{W}_1, \cdots, \overline{W}_{\frac{M}{\gcd(K,M)}-1}\} 
%\end{eqnarray}
%Note that $K_1 + N_2 M = K_1 + K_2 = K$ so that all messages are stored.

For instance, consider the setting in Example 1, where $M=3, K=7, N=4$. Then $N_2 = 1, N_1 = 3, K_2 = 3, K_1 = 4$ and the storage design is as follows.
\begin{eqnarray}
S_1 &=& \{W_1, W_2, W_4\} \\
S_2 &=& \{W_1, W_3, W_4\} \\
S_3 &=& \{W_2, W_3 \} \\
S_4 &=& \{W_5, W_6, W_7\} 
\end{eqnarray}

Suppose $W_k, k \in [1:K]$ is desired. In the linear delivery scheme, each answer from the first server set has $D_i = 1, \forall i \in [1:N_1]$ symbol, and each answer from the second server set has $D_i = L, \forall i \in [N_1 + 1: N]$ symbols. Then the rate achieved is 
$R =\frac{L}{N_1 + L N_2} = \frac{l}{N + (l-1)N_2}$, as desired (refer to (\ref{eq:ldef})). 

The answers are shown below.
\begin{eqnarray}
{A}_i^{[k]} &=& {\bf F}_i^{[k]} W_k + {\bf H}_i Z
\end{eqnarray}
where
\begin{eqnarray}
\begin{array}{lc}
 \mbox{if}~{i} \in [1: N_1], & ~\mbox{${\bf F}^{[k]}_{i}$ is a $1\times L$ vector, and ${\bf H}_i$ is a $1\times(D_\Sigma - L)$ vector},~ \\
 \mbox{else}~{i} \in [N_1 + 1: {N}], & ~\mbox{${\bf F}^{[k]}_{i}$ is an $L\times L$ matrix, and ${\bf H}_i$ is an $L\times(D_\Sigma - L)$ matrix}.~
\end{array}
\end{eqnarray}
Then the collection of all answers are as follows.
\begin{eqnarray}
{\bf A}^{[k]} &=& {\bf F}^{[k]}_{D_\Sigma \times L} W_k + {\bf H}_{D_\Sigma \times (D_\Sigma-L)} Z, \\
\mbox{where} ~~~~ {\bf F}^{[k]} &=& [{\bf F}_1^{[k]}; {\bf F}_2^{[k]}; \cdots; {\bf F}_N^{[k]}], {\bf H} = [{\bf H}_1; {\bf H}_2; \cdots; {\bf H}_N].
\end{eqnarray}

The decoding filtering matrix is denoted by ${\bf G}_{L \times D_\Sigma}$. Then we have
\begin{eqnarray}
W_k = 
{\bf G} {\bf A}^{[k]} &=& {\bf G} {\bf F}^{[k]} W_k + {\bf G} {\bf H} Z, \\
\Rightarrow ~~~~~~ {\bf G} {\bf F}^{[k]} &=& {\bf I}_{L} \Rightarrow 
\begin{array}{ll} 
\mbox{If}~{k} \in [1: K_1], & {\bf G}_{[:, \overrightarrow{\mathcal{N}}_k]} {\bf F}^{[k]}_{[\overrightarrow{\mathcal{N}}_k, :]} = {\bf I}_L, \\
 \mbox{else}~{k} \in [K_1 + 1: {K}], & {\bf G}_{[:, N_1+ ({\mathcal{N}}_k -N_1-1)L + 1: N_1 + ({\mathcal{N}}_k -N_1) L]} {\bf F}^{[k]}_{{\mathcal{N}}_k} = {\bf I}_L,
\end{array} 
\label{eq:gcon1k} \\
{\bf G} {\bf H} &=& {\bf 0}_{L \times (D_\Sigma-L)} \label{eq:gcon2k}
\end{eqnarray}
and all other unspecified sub-matrices of ${\bf F}^{[k]}$ are zero matrices, due to the storage constraint.

To satisfy (\ref{eq:gcon1k}), (\ref{eq:gcon2k}), we set ${\bf G}, {\bf F}^{[1]}, {\bf F}^{[2]}, \cdots, {\bf F}^{[K]}, {\bf H}$ as follows.
\begin{eqnarray}
&& {\bf G} = \left[
\begin{array}{ccccc}
{\bf I}_{L} & {\bf V}_{L \times (D_{\Sigma}-L)}
\end{array}
\right], ~\mbox{where ${\bf V}$ is a Cauchy matrix such that $V_{ij} = \frac{1}{\alpha_i - \beta_j}, \alpha_i \neq \beta_j$}, \notag\\.
&&{\bf H} = \left[
\begin{array}{cc}
{\bf V}_{L \times (D_\Sigma-L)};
-{\bf I}_{D_\Sigma-L}
\end{array}
\right], \\
&& \begin{array}{ll}
{\bf F}^{[k]}_{[\overrightarrow{\mathcal{N}}_k, :]} = {\bf G}_{[:, \overrightarrow{\mathcal{N}}_k]}^{-1}, & \mbox{if}~\overline{k} \in [1: K_1], \\
{\bf F}^{[k]}_{{\mathcal{N}}_k} = {\bf G}_{[:, N_1+ ({\mathcal{N}}_k -N_1 -1)L + 1: N_1 + ({\mathcal{N}}_k - N_1) L]}^{-1},  &  \mbox{else}~\overline{k} \in [K_1 + 1: {K}].
\end{array}\label{eq:ninvk}
\end{eqnarray}
By the same reasoning as that in Section \ref{sec:general}, the matrices in (\ref{eq:ninvk}) have full rank so that their inverse matrices are well defined.
Now correctness constraints are satisfied, and privacy is guaranteed by the observation that 
\begin{eqnarray}
&& H({\bf A}^{[k]}) = D_\Sigma = H(W_k, Z) \\
\Longleftrightarrow && {\bf B}^{[k]}_{D_{\Sigma} \times {D_\Sigma}} = [{\bf F}^{[k]} ~~{\bf H}], ~\forall k \in [1:K] ~\mbox{have full rank}.
\end{eqnarray}
The proof for ${\bf B}^{[k]}$ being full rank follows similarly from that in Section \ref{sec:general} and the details are thus omitted. %The proof is thus complete.

\section{Optimality of Achievable Schemes for Examples 1 and 2}
We present the proof for Example 2 first because it is simpler. The proof idea for both examples is the same - we consider all possible storage strategies and show that none of them may outperform the achieved rate. For each storage strategy, we argue that certain combinatoric structure must exist and the structure leads to a rate upper bound (using Lemma \ref{lemma:basic})\footnote{Our proof is brute-force based in essence. This is the reason that we are not able to generalize this converse proof. However, we are not aware of any setting where the best achievable rate given by Theorem \ref{thm:ach} is not optimal.}.

\subsection{Example 2. $M=4, K=5$}\label{sec:ex2}
We have two settings to consider, i.e., $N=3$ and $N=4$.
\subsubsection{$N=3$: Proof of $R \leq 2/3$}
%First, consider the setting where $N=3$, we show that $R \leq 2/3$.
We assume without loss of generality that each server stores $M=4$ distinct messages (because storing more messages does not hurt). Then we have $K=5$ messages and $MN = 12$ messages are stored across all servers. %Further, by Lemma \ref{lemma:no_repl}, we may assume that the storage at each server is distinct.
Denote $\mathcal{N}_k, k \in [1:5]$ as the set of servers where $W_k$ is stored, so that $|\mathcal{N}_k|$ represents the number of servers where $W_k$ is stored. We assume that $|\mathcal{N}_1| \geq |\mathcal{N}_2| \geq \cdots \geq |\mathcal{N}_5|$. Therefore, we have a partition of the total storage of $12$ messages.
\begin{eqnarray}
12 = |\mathcal{N}_1| + |\mathcal{N}_2| + |\mathcal{N}_3| + |\mathcal{N}_4| + |\mathcal{N}_5|, ~~~ |\mathcal{N}_k| \in [1:3].
\end{eqnarray}
Note that $|\mathcal{N}_k| \geq 1$ because all messages must be stored somewhere and $|\mathcal{N}_k| \leq 3$ because we only have $N=3$ servers. Because of the range of $|\mathcal{N}_k|$ and the assumption of the monotonic non-increasing property on the $\mathcal{N}_k$ sequence, we only have the following 2 cases.
\begin{eqnarray}
\mbox{Case 1:} && (|\mathcal{N}_1|, |\mathcal{N}_2|, |\mathcal{N}_3|, |\mathcal{N}_4|, |\mathcal{N}_5|) = (3,3,3,2,1) \\
\mbox{Case 2:} && (|\mathcal{N}_1|, |\mathcal{N}_2|, |\mathcal{N}_3|, |\mathcal{N}_4|, |\mathcal{N}_5|) = (3,3,2,2,2)
\end{eqnarray} 
For both cases, the storage design is deterministic (up to permutations of the servers). Denote $(\pi_1, \pi_2, \pi_3)$ as a permutation of the 3 servers $(1,2,3)$. For Case 1, we have
\begin{eqnarray}
S_{\pi_1} &=& \{W_1, W_2, W_3, W_4\} \\
S_{\pi_2} &=& \{W_1, W_2, W_3, W_4\} \\
S_{\pi_3} &=& \{W_1, W_2, W_3, W_5\}
\end{eqnarray}
Using Lemma \ref{lemma:basic} for $W_4$ and $W_5$, we have
\begin{eqnarray}
D_{\pi_1} + D_{\pi_2} &\geq& L \label{eq:ee1} \\
D_{\pi_3} &\geq& L \label{eq:ee2} \\
(\ref{eq:ee1}) + (\ref{eq:ee2}) \Rightarrow D_1 + D_2 + D_3 &\geq& 2L \\
\Rightarrow R = \frac{L}{D_1 + D_2 + D_3} &\leq& 1/2 < 2/3
\end{eqnarray}
For Case 2, we have
\begin{eqnarray}
S_{\pi_1} &=& \{W_1, W_2, W_3, W_4\} \\
S_{\pi_2} &=& \{W_1, W_2, W_3, W_5\} \\
S_{\pi_3} &=& \{W_1, W_2, W_4, W_5\}
\end{eqnarray}
Using Lemma \ref{lemma:basic} for $W_3$, $W_4$ and $W_5$, we have
\begin{eqnarray}
D_{\pi_1} + D_{\pi_2} &\geq& L \label{eq:ee3}\\
D_{\pi_3} + D_{\pi_1} &\geq& L \label{eq:ee4}\\
D_{\pi_2} + D_{\pi_3} &\geq& L \label{eq:ee5}\\
(\ref{eq:ee3}) + (\ref{eq:ee4}) + (\ref{eq:ee5}) \Rightarrow 2(D_1 + D_2 + D_3) &\geq& 3L \\
\Rightarrow R = \frac{L}{D_1 + D_2 + D_3} &\leq& 2/3
\end{eqnarray}
Therefore, for both cases, the rate can not be higher than $2/3$ so that the achieved rate of $R = 2/3$ is optimal.

\subsubsection{$N=4$: Proof of $R \leq 3/4$}
%Second, consider the setting where $N=4$, we show that $R \leq 3/4$.
The proof idea is similar. We consider a partition of the total storage of $MN = 16$ messages to the $K = 5$ messages.
\begin{eqnarray}
16 = |\mathcal{N}_1| + |\mathcal{N}_2| + |\mathcal{N}_3| + |\mathcal{N}_4| + |\mathcal{N}_5|, ~~~|\mathcal{N}_1|\geq \cdots \geq |\mathcal{N}_5|, |\mathcal{N}_k| \in [1:4], \forall k \in [1:5].
\end{eqnarray}
For the partition, we have the following 4 cases.
\begin{enumerate}
\item $(|\mathcal{N}_1|, |\mathcal{N}_2|, |\mathcal{N}_3|, |\mathcal{N}_4|, |\mathcal{N}_5|) = (4,4,4,3,1)$.

In this case, $W_4$ and $W_5$ are stored over 2 disjoint sets of servers. Using Lemma \ref{lemma:basic}, we have $\sum_{i=1}^4 D_i \geq 2L$ so that $R \leq 1/2 < 3/4$.
\item $(|\mathcal{N}_1|, |\mathcal{N}_2|, |\mathcal{N}_3|, |\mathcal{N}_4|, |\mathcal{N}_5|) = (4,4,4,2,2)$.

This case is similar to that above, where $W_4$ and $W_5$ are stored over 2 disjoint sets of servers. Then $R \leq 1/2$ follows.

\item $(|\mathcal{N}_1|, |\mathcal{N}_2|, |\mathcal{N}_3|, |\mathcal{N}_4|, |\mathcal{N}_5|) = (4,4,3,3,2)$.

The storage design is deterministic (up to permutation of the servers). Denote $(\pi_1, \pi_2, \pi_3, \pi_4)$ as a permutation of the 4 servers $(1,2,3,4)$. We have
\begin{eqnarray}
S_{\pi_1} &=& \{W_1, W_2, W_3, W_4\} \\
S_{\pi_2} &=& \{W_1, W_2, W_3, W_4\} \\
S_{\pi_3} &=& \{W_1, W_2, W_3, W_5\} \\
S_{\pi_4} &=& \{W_1, W_2, W_4, W_5\}
\end{eqnarray}
Using Lemma \ref{lemma:basic} for $W_3$, $W_4$ and $W_5$, we have
\begin{eqnarray}
D_{\pi_1} + D_{\pi_2} + D_{\pi_3} &\geq& L \\
D_{\pi_4} + D_{\pi_1} + D_{\pi_2} &\geq& L \\
D_{\pi_3} + D_{\pi_4} &\geq& L \\
\Rightarrow 2(D_1 + D_2 + D_3+D_4) &\geq& 3L \\
\Rightarrow R = \frac{L}{D_1 + D_2 + D_3+D_4} &\leq& 2/3 < 3/4
\end{eqnarray}

\item $(|\mathcal{N}_1|, |\mathcal{N}_2|, |\mathcal{N}_3|, |\mathcal{N}_4|, |\mathcal{N}_5|) = (4,3,3,3,3)$.
The storage is also deterministic. We have
\begin{eqnarray}
S_{\pi_1} &=& \{W_1, W_2, W_3, W_4\} \\
S_{\pi_2} &=& \{W_1, W_2, W_3, W_5\} \\
S_{\pi_3} &=& \{W_1, W_2, W_4, W_5\} \\
S_{\pi_4} &=& \{W_1, W_3, W_4, W_5\}
\end{eqnarray}
Using Lemma \ref{lemma:basic} for $W_2$, $W_3$, $W_4$ and $W_5$, we have
\begin{eqnarray}
D_{\pi_{i_1}} + D_{\pi_{i_2}} + D_{\pi_{i_3}} &\geq& L, ~~~\forall ~\mbox{distinct}~i_1, i_2, i_3 \in [1:4] \\
\Rightarrow D_1 + D_2 + D_3+D_4 &\geq& 4L/3 \\
\Rightarrow R = \frac{L}{D_1 + D_2 + D_3+D_4} &\leq& 3/4
\end{eqnarray}
\end{enumerate}

All cases are covered and we always have $R \leq 3/4$. The proof is thus complete.

\subsection{Example 1. $M=3, K=7, N=4$ and Proof of $R\leq 2/5$}\label{sec:ex1}
We follow the same proof idea presented in the previous section for Example 2.

Consider a partition of the total storage of $MN = 12$ messages to the $K = 7$ messages.
\begin{eqnarray}
16 = |\mathcal{N}_1| + |\mathcal{N}_2| + \cdots + |\mathcal{N}_7|, ~~~|\mathcal{N}_1|\geq \cdots \geq |\mathcal{N}_7|, |\mathcal{N}_k| \in [1:4], \forall k \in [1:7].
\end{eqnarray}
For the partition, we have the following 5 cases.
\begin{enumerate}
\item $(|\mathcal{N}_1|, |\mathcal{N}_2|, \cdots, |\mathcal{N}_7|) = (4,3,1,1,1,1,1)$.

In this case, 4 out of the 5 messages $W_3, W_4, W_5, W_6, W_7$ (each appeares once) are stored over 4 disjoint sets of servers. Using Lemma \ref{lemma:basic}, we have $\sum_{i=1}^4 D_i \geq 4L$ so that $R \leq 1/4 < 2/5$.

\item $(|\mathcal{N}_1|, |\mathcal{N}_2|, \cdots, |\mathcal{N}_7|) = (4,2,2,1,1,1,1)$.

In this case, 3 out of the 6 messages $W_2, W_3, W_4, W_5, W_6, W_7$ are stored over 3 disjoint sets of servers. Using Lemma \ref{lemma:basic}, we have $\sum_{i=1}^4 D_i \geq 3L$ so that $R \leq 1/3 < 2/5$.

\item $(|\mathcal{N}_1|, |\mathcal{N}_2|, \cdots, |\mathcal{N}_7|) = (3,3,2,1,1,1,1)$.

Consider the last 4 messages $W_4, W_5, W_6, W_7$ (each appears once). If these 4 messages appear in 3 servers, then similar as the case above, we have $R \leq 1/3$. Henceforth, we focus on the setting where these 4 messages appear in 2 servers. The allocation of these 4 messages to the 2 servers might be $3+1$ or $2+2$. It is easy to see that for both settings, $W_3$ must appear in the other 2 remaining servers so that we have 3 messages that appear in 3 disjoint sets of servers, i.e., $R\leq 1/3$.

\item $(|\mathcal{N}_1|, |\mathcal{N}_2|, \cdots, |\mathcal{N}_7|) = (3,2,2,2,1,1,1)$.

Consider the last 3 messages $W_5, W_6, W_7$ (each appears once). If these 3 messages appear in 3 servers, then similar as the case above, we have $R \leq 1/3$. Henceforth, we focus on the setting where these 3 messages appear in 1 server or 2 servers. 

When $W_5, W_6, W_7$ appear in 1 server, the storage is deterministic. We have
\begin{eqnarray}
S_{\pi_1} &=& \{W_1, W_2, W_3\} \\
S_{\pi_2} &=& \{W_1, W_2, W_4\} \\
S_{\pi_3} &=& \{W_1, W_3, W_4\} \\
S_{\pi_4} &=& \{W_5, W_6, W_7\}
\end{eqnarray}
Using Lemma \ref{lemma:basic} for $W_2$, $W_3$, $W_4$ and $W_5$, we have
\begin{eqnarray}
D_{\pi_{i_1}} + D_{\pi_{i_2}} &\geq& L, ~~~\forall ~\mbox{distinct}~i_1, i_2  \in [1:3] \\
D_{\pi_4} &\geq& L, \\
\Rightarrow \sum_{i=1}^4 D_i &\geq& 5L/2, ~~R = \frac{L}{\sum_{i=1}^4 D_i} \leq 2/5
\end{eqnarray}

When $W_5, W_6, W_7$ appear in 2 servers, we have
\begin{eqnarray}
S_{\pi_1} &=& \{\times, \times, \times\} \\
S_{\pi_2} &=& \{\times, \times, \times \} \\
S_{\pi_3} &=& \{\times, \times, W_5\} \\
S_{\pi_4} &=& \{\times, W_6, W_7\}
\end{eqnarray}
where $\times$ represents place-holders for the remaining messages, $W_1$ (will appear 3 times), $W_2, W_3, W_4$ (will appear 2 times each). By enumerating all possibilities, it is easy to see that there exists 1 message out of $W_2, W_3, W_4$ that appears only in $S_{\pi_1}, S_{\pi_2}$. Combining this message with $W_5, W_6$, we have 3 messages that appear in 3 disjoint sets of servers and it follows that $R \leq 1/3$.

\item $(|\mathcal{N}_1|, |\mathcal{N}_2|, \cdots, |\mathcal{N}_7|) = (2,2,2,2,2,1,1)$.

We have 2 possibilities here, depending on how many servers will be occupied by the last 2 messages.

When $W_6, W_7$ (each appears once) appear in 2 servers, there must exist 1 message out of $W_1, W_2, W_3, W_4, W_5$ that appears only in the 2 remaining servers. Similarly, we have 3 messages that appear in 3 disjoint sets of servers and $R \leq 1/3$.

When $W_6, W_7$ (each appears once) appear in 1 server, we have (denote $(\gamma_1, \cdots, \gamma_5)$ as a permutation of $(1,\cdots,5)$)
\begin{eqnarray}
S_{\pi_1} &=& \{\times, \times, \times\} \\
S_{\pi_2} &=& \{\times, \times, \times \} \\
S_{\pi_3} &=& \{\times, \times, W_{\gamma_5}\} \\
S_{\pi_4} &=& \{W_{\gamma_5}, W_6, W_7\}
\end{eqnarray}
where $\times$ represents place-holders for $W_{\gamma_1}, \cdots, W_{\gamma_4}$ (each appears twice). By a similar reasoning as that in the above case, we must have 1 message out of $W_{\gamma_1}, \cdots, W_{\gamma_4}$ that appears only in $S_{\pi_1}, S_{\pi_2}$. Therefore, 3 messages appear in 3 disjoint sets of servers and $R \leq 1/3$.
\end{enumerate}

To summarize, no matter how we design the storage, the rate is always bounded above by $2/5$ so that the proof is complete.

\section{Discussion}
Motivated by dataset privacy, we introduce the problem of private information delivery, where one out of $K$ messages is sent from a set of servers to a user while the delivered message index remains a secret. We take an information theoretic approach to this problem and adopt the capacity as the performance metric (parallel to the recent line of private information retrieval \cite{PIRfirst, Sun_Jafar_PIR, Sun_Jafar_TPIR, Banawan_Ulukus}, where the privacy of the user is considered).
We propose information theoretic converses that capture this privacy constraint and vector linear coding schemes that satisfy perfect privacy. The rate upper and lower bounds are tight for a wide range of system parameters. We consider the elemental model where the messages are replicated, the user behaves nicely, and a single message is delivered, leaving much room for generalizations. %This work represents a step towards understanding privacy using information theoretic tools.

We have focused exclusively on the metric of rate while the amount of randomness is ignored. The interplay between the communicate rate and the randomness size is an interesting future direction. Further, we are taking a coarse look at the randomness as we assume the same random variable is shared by all servers. It is not hard to see that this is not necessary and we only need the randomness variables to be correlated. A finer view on the rate region of the correlated randomness variables (instead of the sum randomness rate) will shed light on the consumption of randomness. %This research avenue might be technically challenging. For example, although the transformation in Lemma \ref{lemma:no_repl} preserves the rate, the randomness consumption might change. Also, when considering the rate converse, we do not use the fact that the same decoding mapping must be used for all messages (which is crucial for the randomness size converse \cite{Sun_Anonymous}).

%Randomness. for example, lemma XX does not work
%(same decoding mapping is not used)
%Finer view of individual and joint randomness etc

Finally, we mention the connection of the private information delivery problem to the anonymous communications problem \cite{Sun_Anonymous, Peng, Ren_Wu, Danezis_Diaz}, where the identity that needs to be hidden is the transmitters, receivers and their associations. Under many circumstances (e.g., \cite{Sun_Anonymous}), the identity of the delivered message in private information delivery is intimately related to the identity of the nodes in an anonymous communication network. As a result, it is interesting to explore the implications and extensions of the techniques in this work to  anonymous communication networks.

%\section*{Appendix: Symmetrization}

\let\url\nolinkurl
\bibliographystyle{IEEEtran}
\bibliography{Thesis}
\end{document}